\documentclass[]{aa}
\usepackage{calc}
\usepackage{amsmath}
\usepackage{subcaption} 
\usepackage{float}
\usepackage{placeins}
\usepackage{graphicx}
\usepackage{mathtools}
\usepackage{tabularx} 

\usepackage{url}
\usepackage[normalem]{ulem}
\usepackage{soul}

\DeclareMathAlphabet{\mathlib}{OT1}{LinuxLibertineT-OsF}{m}{it}
\DeclareMathAlphabet{\mathbio}{OT1}{LinuxBiolinumT-OsF}{m}{it}
\mathcode`*=\string"8000
\begingroup
\catcode`*=\active
\xdef*{\noexpand\textup{\string*}}
\endgroup

\usepackage{natbib}
\bibpunct{(}{)}{;}{a}{}{,} 
\usepackage[colorlinks,allcolors=blue]{hyperref}

\def\H0{{\rm \,km\,s^{-1}\,Mpc^{-1}}}

\begin{document} 

\title{Sulphur and carbon isotopes towards Galactic centre clouds}

          \author{P. K. Humire, 
          \inst{1}
          V. Thiel,
          \inst{1}
          C. Henkel,
          \inst{1,2,3}
          A. Belloche,
          \inst{1}
          J.-C. Loison,
          \inst{4}
		  T. Pillai,
          \inst{1,5}
          D. Riquelme,
          \inst{1}
          V. Wakelam,
          \inst{6}
          N. Langer,
          \inst{1,7}
          A. Hern\'andez-G\'omez,
          \inst{1}
          R. Mauersberger,
          \inst{1}
          \and
          K. M. Menten
          \inst{1}
          }


\authorrunning{Humire et al. 2019}
\titlerunning{Sulphur and Carbon isotopes towards Sgr. A}

          \institute{Max-Planck-Institut f\"ur Radioastronomie, Auf dem H\"ugel 69, 53121 Bonn, Germany\\
              \email{phumire@mpifr-bonn.mpg.de}
          \and      
              Dept. of Astron., King Abdulaziz University, P.O. Box 80203, Jeddah 21589, Saudi Arabia
          \and
              Xinjiang Astronomical Observatory, Chinese Academy of Sciences, 830011 Urumqi, China
          \and
                Institut des Sciences Mol\'eculaires (ISM), CNRS, Univ. Bordeaux, 351 cours de la Lib\'eration, F-33400, Talence, France
          \and
                Institute for Astrophysical Research, 725 Commonwealth Ave, Boston University Boston, MA 02215, USA
          \and
                Laboratoire d'astrophysique de Bordeaux, CNRS, Univ. Bordeaux, B18N, all\'ee Geoffroy Saint-Hilaire, F-33615 Pessac, France
          \and  Argelander-Institut f\"ur Astronomie, Universit\"at Bonn, Auf dem H\"ugel 71, 53121 Bonn, Germany
                }
   \date{Received xxx, yyy; accepted xxx, yyy}

\abstract {Measuring isotopic ratios is a sensitive technique used to obtain information on stellar nucleosynthesis and chemical evolution.} {We present measurements of the carbon and sulphur abundances in the interstellar medium of the central region of our Galaxy. The selected targets are the $+$50km\,s$^{-1}$\,Cloud and several line-of-sight clouds towards Sgr\,B2(N).} 
{Towards the $+$50\,km\,s$^{-1}$\,Cloud, we observed the \textit{J}=2--1 rotational transitions of $^{12}$C$^{32}$S, $^{12}$C$^{34}$S, $^{13}$C$^{32}$S, $^{12}$C$^{33}$S, and $^{13}$C$^{34}$S, and the \textit{J}=3--2 transitions of $^{12}$C$^{32}$S and $^{12}$C$^{34}$S with the IRAM-30\,m telescope, as well as the \textit{J}=6--5 transitions of $^{12}$C$^{34}$S and $^{13}$C$^{32}$S with the APEX 12\,m telescope, all in emission. The \textit{J}=2--1 rotational transitions of $^{12}$C$^{32}$S, $^{12}$C$^{34}$S, $^{13}$C$^{32}$S, and $^{13}$C$^{34}$S were observed with ALMA in the envelope of Sgr\,B2(N), with those of $^{12}$C$^{32}$S and $^{12}$C$^{34}$S also observed in the line-of-sight clouds towards Sgr\,B2(N), all in absorption.} {In the $+$50\,km\,s$^{-1}$\,Cloud we derive a $^{12}$C/$^{13}$C isotopic ratio of 22.1$^{+3.3}_{-2.4}$, that leads, with the measured $^{13}$C$^{32}$S/$^{12}$C$^{34}$S line intensity ratio, to a $^{32}$S/$^{34}$S ratio of 16.3$^{+3.0}_{-2.4}$. We also derive the $^{32}$S/$^{34}$S isotopic ratio more directly from the two isotopologues $^{13}$C$^{32}$S and $^{13}$C$^{34}$S, which leads to an independent $^{32}$S/$^{34}$S estimation of 16.3$^{+2.1}_{-1.7}$ and 17.9$\pm$5.0 for the $+$50\,km\,s$^{-1}$\,Cloud and Sgr\,B2(N), respectively. We also obtain a $^{34}$S/$^{33}$S ratio of 4.3$\pm$0.2 in the $+$50\,km\,s$^{-1}$\,Cloud.}
{Previous studies observed a decreasing trend in the $^{32}$S/$^{34}$S isotopic ratios when approaching the Galactic centre. Our result indicates a termination of this tendency at least at a galactocentric distance of 130$_{-30}^{+60}$\,pc. This is at variance with findings based on $^{12}$C/$^{13}$C, $^{14}$N/$^{15}$N and $^{18}$O/$^{17}$O isotope ratios, where the above-mentioned trend is observed to continue right to the central molecular zone. This can indicate a drop in the production of massive stars at the Galactic centre, in the same line as recent metallicity gradient ([Fe/H]) studies, and opens the work towards a comparison with Galactic and stellar evolution models.}

   \keywords{Galaxy: abundances -- Galaxy: centre -- Galaxy: evolution --
ISM: abundances -- ISM: molecules -- radio lines: ISM  }

   \maketitle
   
%

\section{Introduction}

\label{introduction}
Studying stellar nucleosynthesis and chemical enrichment of rare isotopes of a given element at optical wavelengths is difficult because the observed atomic isotope lines are usually affected by blending  \citep[e.g.][]{Hawkins1987,Levshakov2006,Ritchey2011}. However, at radio and (sub)millimeter wavelengths, transitions from rare isotopic substitutions of a given molecular species, called isotopologues, are well separated in frequency from their main species, typically by a few percent of their rest frequency.

While the relative abundances of C, N, and O isotopes provide information on carbon–nitrogen–oxygen (CNO) and helium burning, sulphur isotopes allow us to probe late evolutionary stages of massive stars and supernovae (SNe) of Type Ib/c and II (oxygen-burning, neon-burning, and s-process nucleosynthesis) \citep{Wilson1994, Chin1996, Mauersberger1996}, filling a basic gap in our understanding of stellar nucleosynthesis and the chemical evolution of the universe \citep[e.g.][]{Wang2013}.

In the interstellar medium (ISM), atomic sulphur is thought to freeze out on dust grain mantles and to be later released from the grains due to shocks, leading to the formation of several molecular species in the gas phase, such as OCS, SO$_{2}$, H$_{2}$S, and H$_{2}$CS \citep{Millar1990}, which serve as both shock and high-mass star formation tracers in starburst galaxies \citep{Bayet2008}. 

Among the sulphur-bearing compounds, CS (carbon monosulfide) is the most accessible molecular species: its lines are ubiquitous in the dense ISM and tend to be strong at sites of massive star formation in the spiral arms of our Galaxy, in the Galactic centre (GC) region and in external galaxies \citep[e.g.][]{Linke1980,Mauersberger1989,Bayet2009,Kelly2015}.

Sulphur has four stable isotopes: $^{32}$S, $^{33}$S, $^{34}$S, and $^{36}$S. Their solar system fractions are 95.018:0.750:4.215:0.017 \citep{Lodders2003}, respectively. In the ISM, \citet{Chin1996} found a relation between $^{32}$S/$^{34}$S isotope ratios and their galactocentric distance ($D$\textsubscript{GC}) of $^{32}$S/$^{34}$S=(3.3$\pm$0.5)($D$\textsubscript{GC}/kpc)$+$(4.1$\pm$3.1) by using a linear least-squares fit to the unweighted data, with a correlation coefficient of 0.84, while no correlation was obtained between $^{34}$S/$^{33}$S ratios and $D$\textsubscript{GC}. However, most of the sources observed in that study are located within the galactocentric distance range 5.5\,$\leq$\,$D\rm{_{GC}} \leq$\,7.0 kpc, with the minimum distance at 2.9\,kpc from the Galactic centre. Therefore, it is important to also cover the inner region of the Milky Way to find out whether the trend proposed by \citet{Chin1996} is also valid for the inner Galaxy as has been reported for the $^{12}$C/$^{13}$C \cite[see e.g.][]{Yan2019}, $^{14}$N/$^{15}$N \citep{Adande2012} and $^{18}$O/$^{17}$O \citep{Wouterloot2008,Zhang2015} isotopic ratios. 

The GC region harbours one of the most intense and luminous sites of massive star formation in the Galaxy, Sgr\,B2 \citep{Molinari2014,Ginsburg2018}. It provides an extreme environment in terms of pressure, turbulent Mach number, and gas temperature \citep{Ginsburg2016} over a much more extended region than encountered in star-forming regions throughout the Galactic disc \citep{Morris1996,Ginsburg2016,Schwoerer2019,Dale2019}. These conditions are comparable to those in starburst galaxies \citep{Belloche2013,Schwoerer2019}. We therefore expect unique results in this GC study from sulphur ratios, which are a tool for tracing stellar processing (see Sect.\,\ref{discussion}). For a compilation of sulphur ratios determined in our Milky Way, we refer to Tables 2 and 7 in \citet{Mauersberger2004} and \citet{Muller2006}, respectively.

As is true for our Galaxy, detections of $^{34}$S in extragalactic objects remain scarce. Some observations, also accounting for $^{12}$C/$^{13}$C ratios (using the double-isotope ratio method, Sect.\,\ref{doubleisotopemethod}) led to values of $\sim$16--25 and 13.5$\pm$2.5 for the $^{32}$S/$^{34}$S ratio in the nuclear starbursts of NGC\,253 and NGC\,4945, respectively \citep{Wang2004,Henkel2014}, although the value for NGC\,4945 might be underestimated due to saturation of the CS lines \citep{Martin2010}. A ratio of 20$\pm$5 was obtained for N159 in the Large Magellanic Cloud \citep{Johansson1994}. At redshift $z$=0.89, a $^{32}$S/$^{34}$S ratio of 10$\pm$1 has been derived using absorption lines from the spiral arm of a galaxy located along the line of sight (l.o.s.) towards a radio loud quasar \citep{Muller2006}.
 
In the present study we focus on the \textit{J}=2--1 transitions of $^{12}$C$^{32}$S (hereafter CS), $^{12}$C$^{34}$S (hereafter C$^{34}$S), $^{13}$C$^{32}$S (hereafter $^{13}$CS), $^{12}$C$^{33}$S (hereafter C$^{33}$S), and $^{13}$C$^{34}$S, the \textit{J}=3--2 transitions of CS and C$^{34}$S, and the \textit{J}=6--5 transitions of C$^{34}$S and $^{13}$CS, all observed together towards the Sgr\,A Complex \citep[see, e.g.,][]{Sandqvist2015}. We have also studied absorption features caused by the envelope of Sgr\,B2(N) in the \textit{J}=2--1 rotational transition of CS, C$^{34}$S, $^{13}$CS and $^{13}$C$^{34}$S, as well as CS and C$^{34}$S absorption features caused by l.o.s. clouds towards Sgr\,B2(N). These absorption and emission profiles allow us to obtain $^{12}$C/$^{13}$C and the missing $^{32}$S/$^{34}$S ratios close to the Galactic nucleus. Expanding the database for sulphur isotope ratios in the GC region is important in order to constrain models of stellar interiors as well as models of the chemical evolution of the Galaxy \citep[e.g.][]{Kobayashi2011}.

The paper is organised as follows. In Sect.\,\ref{observations} we describe the observations. In Sect.\,\ref{sources} we describe in detail our targets. In Sect.\,\ref{results} we present measured opacities and isotopologue ratios from CS species, the modelling of our Sgr\,B2(N) data, and a comprehensive study of CS fractionation. In Sect.\,\ref{previous studies} we compare our results with previous studies. In Sect.\,\ref{discussion} we discuss the results in the context of trends with galactocentric distance and give some explanations for our findings, before summarising and concluding in Sect.\,\ref{summaryandconclusions}. 

\section{Observations}
\label{observations}

\subsection{\texorpdfstring{$+$50\,km\,s$^{-1}$\,Cloud}{+50 km s Cloud}}

The $+$50\,km\,s$^{-1}$\,Cloud observations were conducted with the IRAM 30\,m and the APEX 12\,m telescopes over a period of 1.5 years, from 2015 May to 2016 September, under varying weather conditions. The observed position was EQ\,J2000: $17^\mathrm{h}45^\mathrm{m}50.20^\mathrm{s},-28^\mathrm{\circ} 59^\mathrm{\arcmin} 41.1^\mathrm{\arcsec}$\, for both telescopes, and the representative spectral resolution was 0.6\,km\,s$^{-1}$. With the IRAM 30\,m telescope, three frequency set-ups were observed with the E090 and E150 receivers in combination with the Fast Fourier Transform Spectrometer (FFTS, at 195\,kHz resolution mode\footnote{http://www.iram.es/IRAMES/mainWiki/Backends}). For the observations presented in this paper, we covered the 93.2--100.98 GHz frequency range (CS, C$^{34}$S, and C$^{33}$S \textit{J}=2--1) in one set-up with the E090 receiver. In a separate E090 set-up, our tuning covered the frequency range 85.5-93.3 GHz ($^{13}$CS and $^{13}$C$^{34}$S \textit{J}=2--1), while simultaneously the E150 receiver covered the frequency range from 143.5 to 151.3 GHz (for CS and C$^{34}$S \textit{J}=3--2). The observations were conducted in total power position switching mode. No spectral contamination was found in our off-source reference position ($17^\mathrm{h}46^\mathrm{m}10.4^\mathrm{s},-29^\mathrm{\circ} 07^\mathrm{\arcmin} 08^\mathrm{\arcsec}$). The main beam efficiencies for our IRAM 30\,m measurements were computed using the Ruze formalism \citep{Ruze1966}. Adopted values were 0.8 and 0.7 at 98 and 147\,GHz, respectively\footnote{calculated following the table in \url{https://www.iram.fr/GENERAL/calls/w08/w08/node20.html}}. We discarded data taken under poor weather conditions (precipitable water vapour content (pwv) $>$ 7\,mm) by discarding data taken with system temperatures $>$ 500\,K. The representative half-power beam widths (HPBW) values are about 25\arcsec\,at 98\,GHz and 17\arcsec\,at 147 GHz for the IRAM 30\,m observations.

The Atacama Pathfinder Experiment 12\,m telescope (APEX) 12\,m \citep{Gusten2006} was used for observations of the \textit{J}=6--5 lines of the CS, $^{13}$CS, and $^{13}$C$^{34}$S  isotopologues at roughly 280 GHz. The measurements were conducted simultaneously using the FLASH345 \citep{Klein2014} receiver connected to the extended FFTS (XFFTS) backend. These observations were also executed in total power position switching mode and the same off-source reference position, which was found to be clean. The HPBW was about 22\arcsec\ at the observed frequency and the adopted main beam efficiency was 0.7. 

All line intensities are reported in main beam brightness temperature units ($\textit{T}_{\textrm{MB}}$). While the spectral resolution was instrument dependent (between 0.4 and 0.6 km\,s$^{-1}$ for the IRAM and 0.08 km\,s$^{-1}$ for the APEX data), all spectra were smoothed to a resolution of 3\,km\,s$^{-1}$ for analysis. 

The data were reduced with the GILDAS package\footnote{\url{https://www.iram.fr/IRAMFR/GILDAS}} and required minimal flagging, followed by a baseline subtraction of order two.

\subsection{Line-of-sight clouds towards Sgr B2(N)}
\label{los}

\begin{figure}
  \includegraphics[,width=10cm, trim = 2.cm 2.cm 7.5cm 1cm, clip=True]{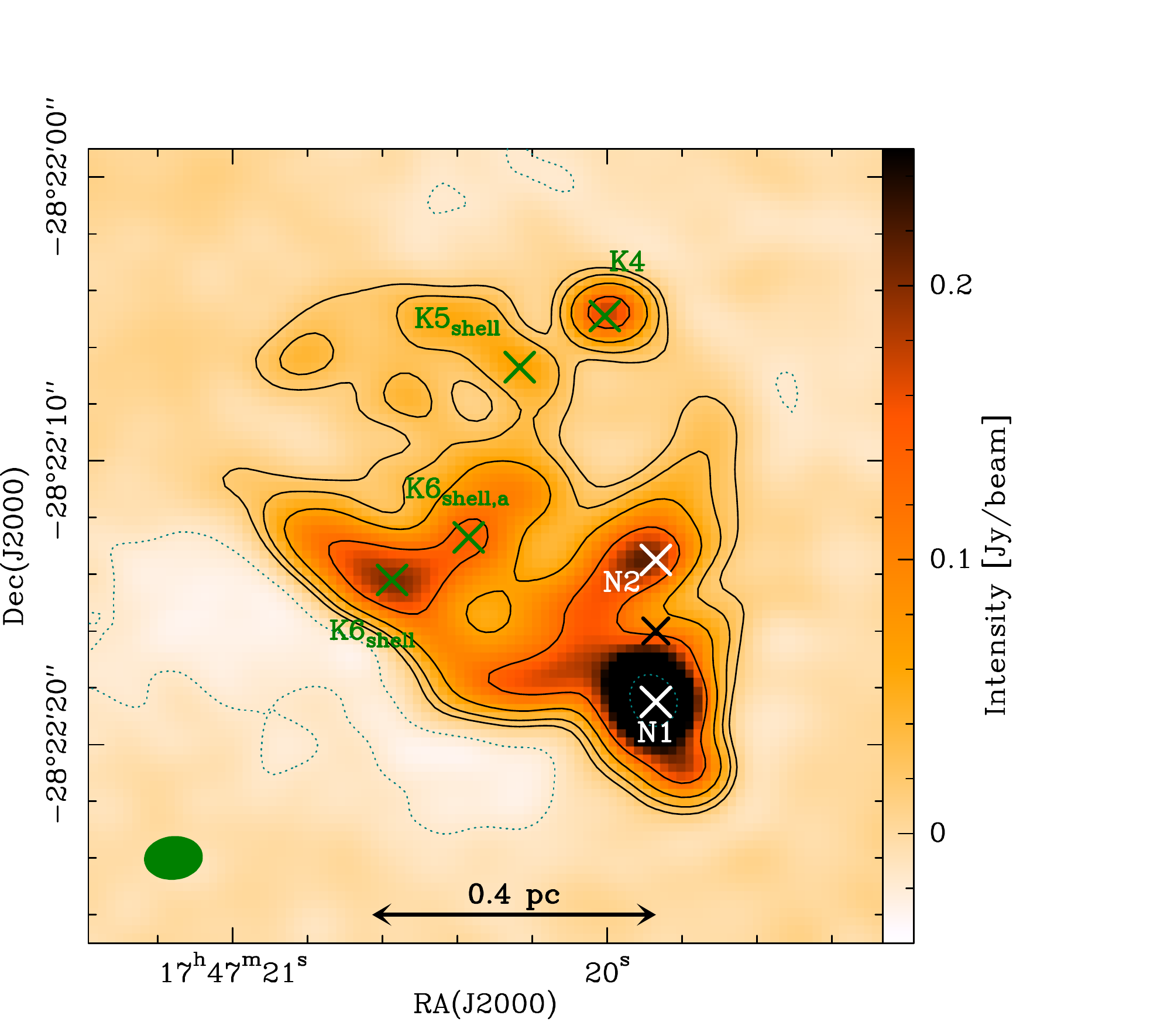}
  \caption{ALMA continuum map of Sgr\,B2(N) at 85\,GHz. The black contour lines show the flux density levels at 3$\sigma$, 6$\sigma$, 12$\sigma$, and 24$\sigma$ and the dotted lines indicate -3$\sigma$, where $\sigma$ is the rms noise level of 5.4\,mJy\,beam$^{-1}$. The white crosses denote the positions of the two main hot cores, Sgr\,B2(N1) and Sgr\,B2(N2). The black cross, located between the white ones, indicates the phase centre. The green crosses show peaks of continuum emission selected for our absorption study. They label the ultra compact H\,{\sc ii} region K4, two peaks in the shell of the H\,{\sc ii} region K6 and one peak in the shell of the H\,{\sc ii} region K5 \citep{Gaume1995}. The green ellipse in the lower left corner shows the size of the synthesised beam of $\sim$1\farcs6 (see Table\,\ref{tab:1}).}

  \label{fig:SgrB2}
\end{figure}

For Sgr\,B2(N), we used the Exploring Molecular Complexity with ALMA (EMoCA) survey \citep{Belloche2016} that was performed with the Atacama Large Millimeter/submillimeter Array (ALMA) in the direction of this source. The centre of the observed field (see Fig.\,\ref{fig:SgrB2}) lies in the middle (EQ\,J2000: $17^\mathrm{h}47^\mathrm{m}19.87^\mathrm{s},-28^\mathrm{\circ} 22^\mathrm{\arcmin} 16^\mathrm{\arcsec}$) between the two main hot cores, N1 and N2, which are separated by 4\farcs9 along the north--south direction, or around 0.19\,pc in projection assuming a distance to the Galactic centre, Sgr\,A$^{\ast}$, of 8.122\,kpc \citep{GravityCollaboration2018}. The survey covers the frequency range from 84.1 to 114.4\,GHz, which includes carbon monosulfide \textit{J}=2--1 lines, with a spectral resolution of 488 kHz (1.7 to 1.3\,km\,s$^{-1}$) at a median angular resolution of 1\farcs6, or $\sim$0.06\,pc. The average noise level is $\sim$3\,mJy\,beam$^{-1}$ per channel. Details of the calibration and deconvolution of the data are reported in \citet[][Sect. 2.2]{Belloche2016}. For this work, we corrected the data for primary beam attenuation. Several isotopologues of CS are detected in the EMoCA survey. To determine the $^{32}$S/$^{34}$S and $^{12}$C/$^{13}$C isotopic ratios, we use four isotopologues: CS, C$^{34}$S, $^{13}$CS, and $^{13}$C$^{34}$S.

A list of the observed transitions of CS isotopologues and some associated parameters is given in Table \ref{tab:1}. 

\begin{table*}[!t]
\caption{Some observational and physical parameters\tablefootmark{a} for the five measured CS isotopologue transitions in the $+$50km\,s$^{-1}$ Cloud and the four transitions from the l.o.s. clouds towards Sgr\,B2(N), which are also located in the Galactic centre region.}
\label{tab:1}      
\centering
\begin{tabular}{l l c c r r r r}
    \hline \hline
        Target       & Telescope    & Isotopologue     &Transition       & \multicolumn{1}{c}{$\nu_{0}$\tablefootmark{b}}    & \multicolumn{1}{c}{$\textit{E}_\mathrm{up}/k$\tablefootmark{c}} & $A_\mathrm{u,l}$\tablefootmark{d}     &\multicolumn{1}{c}{$\mathrm{HPBW}$\tablefootmark{e}} \\ 
                 &              &                  &                 & \multicolumn{1}{c}{[GHz]}        & \multicolumn{1}{c}{[K]}             & [s$^{-1}$]           & \multicolumn{1}{c}{[\arcsec]} \\ \hline \\    
    $+$50\,Cloud  &IRAM\,30\,m     &$^{13}$C$^{34}$S  & 2--1             & 90.9260260                       & 6.55                             & $1.19\times 10^{-5}$ & 27.0\\ 
    $+$50\,Cloud  &IRAM\,30\,m     &$^{13}$CS         & 2--1             & 92.4943080                       & 6.66                             & $1.41\times 10^{-5}$ & 26.6\\ 
    $+$50\,Cloud  &IRAM\,30\,m     &C$^{34}$S         & 2--1             & 96.4129495                       & 6.25                             & $1.60\times 10^{-5}$ & 25.5\\     
    $+$50\,Cloud  &IRAM\,30\,m     &C$^{33}$S         & 2--1             & 97.1720639                       & 7.00                             & $1.63\times 10^{-5}$ & 25.3\\         
    $+$50\,Cloud  &IRAM\,30\,m     &CS                & 2--1             & 97.9809533                       & 7.05                             & $1.67\times 10^{-5}$ & 25.1\\
    $+$50\,Cloud  &IRAM\,30\,m     &C$^{34}$S         & 3--2             & 144.6171007                      & 11.80                            & $5.74\times 10^{-5}$ & 17.0\\     
    $+$50\,Cloud  &IRAM\,30\,m     &CS                & 3--2             & 146.9690287                      & 14.10                            & $6.05\times 10^{-5}$ & 16.6\\ 
    $+$50\,Cloud  &APEX\,12\,m     &$^{13}$CS         & 6--5             & 277.4554050                      & 46.60                            & $4.40\times 10^{-4}$ & 22.1\\ 
    $+$50\,Cloud  &APEX\,12\,m     &C$^{34}$S         & 6--5             & 289.2090684                      & 38.19                            & $4.81\times 10^{-4}$ & 21.2\\ 
    Sgr\,B2(N)    &ALMA            &$^{13}$C$^{34}$S  & 2--1             & 90.9260260                       & 6.55                             & $1.19\times 10^{-5}$ & 1.8$\times$1.6\\ 
    Sgr\,B2(N)    &ALMA            &$^{13}$CS         & 2--1             & 92.4943080                       & 6.66                             & $1.41\times 10^{-5}$ & 2.9$\times$1.5\\
    Sgr\,B2(N)    &ALMA            &C$^{34}$S         & 2--1             & 96.4129549                       & 6.25                             & $1.60\times 10^{-5}$ & 1.9$\times$1.4\\
    Sgr\,B2(N)    &ALMA            &CS                & 2--1             & 97.9809533                       & 7.05                             & $1.67\times 10^{-5}$ & 1.8$\times$1.3\\ 
 \hline 
\end{tabular}
\tablefoot{\tablefoottext{a}{The spectroscopic information for each molecule is taken from the Cologne Database for Molecular Spectroscopy \citep[CDMS,][]{Mueller2005,Endres2016}.} \tablefoottext{b}{Rest frequency.} \tablefoottext{c}{Upper energy level.} \tablefoottext{d}{Einstein coefficient for spontaneous emission from upper $u$ to lower $l$ level.} \tablefoottext{e}{Half-power beam width. For the IRAM sources it was calculated following Eq.\,1 in \url{http://www.iram.es/IRAMES/telescope/telescopeSummary/telescope_summary.html.}}}
\end{table*}

\begin{table*}[!t]
\caption{Line parameters for the nine measured CS isotopologue transitions in the $+$50\,km\,s$^{-1}$\,Cloud.}
\label{tab:2}      
\centering
\begin{tabular}{l c r r r r r}
    \hline \hline
    Isotopologue & Transition & \multicolumn{1}{c}{$\int$\textit{T}$_{\rm{mb}}\rm{dV}$} & \multicolumn{1}{c}{Peak velocity} & \multicolumn{1}{c}{FWHM} & \multicolumn{1}{c}{\textit{T}$_{\rm{mb}}^{\rm{peak}}$} &\multicolumn{1}{c}{$\tau$\tablefootmark{c}} \\
    
     &         & \multicolumn{1}{c}{[K\,km\,s$^{-1}$]}  & \multicolumn{1}{c}{[km\,s$^{-1}$]}& \multicolumn{1}{c}{[km\,s$^{-1}$]} & \multicolumn{1}{c}{[K]} &  \\ \hline  \\
     
    $^{13}$C$^{34}$S\tablefootmark{d}   &  2--1       & 1.25$\pm$0.12    & 48.44$\pm$0.92 & 20.99$\pm$2.24 & 0.06$\pm$0.005 & -\\  
    $^{13}$CS                           &  2--1       & 23.28$\pm$0.28   & 46.94$\pm$0.12 & 22.28$\pm$0.29 & 0.98$\pm$0.01 & 0.08$^{+0.05}_{-0.04}$\\     
    C$^{34}$S                           &  2--1       & 32.61$\pm$0.41   & 46.45$\pm$0.13 & 23.07$\pm$0.32 & 1.33$\pm$0.02 & -\\    
    C$^{33}$S                           &  2--1       &  7.73$\pm$0.31   & 47.05$\pm$0.45 & 27.61$\pm$1.14 & 0.26$\pm$0.01 & -\\  
    CS\tablefootmark{$+$}                 &  2--1       &250.00$\pm$6.66   & 48.10$\pm$0.11 & 27.56$\pm$0.51 & 8.52$\pm$0.16 & 1.9$^{+1.1}_{-0.8}$\\ 
    CS\tablefootmark{$-$}                 &  2--1       &-37.62$\pm$6.13   & 48.09$\pm$0.19 & 10.85$\pm$0.82 &-3.26$\pm$0.32 & -\\ 
    C$^{34}$S                           &  3--2       & 21.19$\pm$0.19   & 45.95$\pm$0.09 & 22.39$\pm$0.22 & 0.89$\pm$0.01 & 0.05--0.15   \\     
    CS\tablefootmark{$+$}                 &  3--2       &250.00$\pm$14.22  & 47.87$\pm$0.10 & 25.59$\pm$0.61 & 9.18$\pm$0.73 & 1.0--2.8\\ 
    CS\tablefootmark{$-$}                 &  3--2       &-78.71$\pm$14.69  & 48.45$\pm$0.12 & 13.97$\pm$0.77 &-5.29$\pm$0.71 & -\\ 
    $^{13}$CS                           &  6--5       & 2.39$\pm$0.12    & 43.94$\pm$0.43 & 18.61$\pm$1.04 & 0.120$\pm$0.005 & -\\ 
    C$^{34}$S                           &  6--5       & 3.32$\pm$0.10    & 44.42$\pm$0.26 & 18.51$\pm$0.62 & 0.169$\pm$0.005 & -\\ 
\hline 
\end{tabular}
\tablefoot{\tablefoottext{c}{Peak opacity}.\tablefoottext{d}{a CH$_3$OCH$_3$ (see Sect.\,\ref{peak_opacities_and_line_intesity_ratios} and Fig.\,\ref{fig:50cloudimage}) contribution was subtracted before performing a single Gaussian fit to this line}.\tablefoottext{$+$}{Positive component (in blue in Fig.\,\ref{fig:50cloudimage})}. \tablefoottext{$-$}{Negative component (in green in Fig.\,\ref{fig:50cloudimage}).}}
\end{table*}

\section{Sources}
\label{sources}
The $+$50\,km\,s$^{-1}$\,Cloud (also known as M--0.02--0.07, although the $+$50\,km\,s$^{-1}$\,Cloud could include additional molecular knots on its positive-longitude side; \citep{Ferriere2012}), observed here with IRAM and APEX, is a giant molecular cloud (GMC) of hook or indented-sphere shape, considered to be one of the most prominent high-mass star formation sites in the GC. It has a mass of $\sim$5$\times$10$^{5}$\,M$_{\odot}$; a density of 10$^{4}$--10$^{5}$\,cm$^{-3}$; a gas temperature of 80--100\,K \citep[from NH$_{3}$, CH$_{3}$CN, and CH$_{3}$CCH;][]{Gusten1985}, $\sim$190\,K \citep[from H$_{2}$CO,][]{Ao2013}, or 410$\pm$10\,K \citep[from NH$_{3}$,][]{Mills2013}; and a dust temperature of 20--30\,K \citep{Sandqvist2008}. The energetics of at least a part of the  $+$50\,km\,s$^{-1}$\,Cloud are influenced by the supernova (SN) remnant Sgr\,A\,East \citep{Ferriere2012,Uehara2019}. In CS emission, its dense core peaks at a l.o.s. distance of 3$\pm$3\,pc relative to the Galactic centre \citep{Ferriere2012}, coincident with 1.2\,mm observations \citep[and references therein]{Vollmer2003}, and peaks at ($\Delta\alpha, \Delta \delta$) $\approx$ (3\farcm0, 1\farcm5) with respect to Sgr\,A$^{\ast}$, corresponding to $\approx$7$\pm$3\,pc to the east along the Galactic plane and 4.5$\pm$3\,pc to the south from the Galactic plane \citep[see the explanation for galactocentric Cartesian coordinates in][and their Table 1]{Ferriere2012}.

With ALMA we have observed Sgr\,B2(N)orth, one of the sites of massive star formation associated with Sgr\,B2, the most massive cloud in our Galaxy. The whole Sgr\,B2 complex has a total mass of 10$^{7}$\,M$_{\odot}$ \citep{Goldsmith1990} and gas temperatures of at least 50\,K \citep[from p--H$_{2}$CO in][their Sect. 5.8]{Ginsburg2016}. Most of the mass in Sgr\,B2(N) ($\sim$73\%) is contained in one single core (AN01) \citep{SanchezMonge2017}. The densities and column densities of the hot cores in Sgr\,B2(N) are high \citep[$>$10$^{6}$\,cm$^{-3}$ and $>$10$^{23}$\,cm$^{2}$;][]{SanchezMonge2017,Bonfand2019}. 

Sgr\,B2(N) is located at a galactocentric distance of 130$_{-30}^{+60}$\,pc from Sgr\,A$^{\ast}$ \citep[from][considering the projected distance of 100\,pc as lower limit]{Reid2009} or $\sim$8\,kpc from us. Its diameter is $\sim$0.8\,pc \citep{Lis1990,Schmiedeke2016}. Several diffuse and translucent clouds are detected in absorption along the l.o.s. to Sgr\,B2(N) \citep[see e.g.][]{Greaves1994,Thiel2019}. The exact locations of the different velocity components of the l.o.s. clouds towards Sgr\,B2(N) are not yet fully constrained. In the case of our observations (see Table\,\ref{isotopic_ratios_emoca}), we can only suggest that the lines with negative velocities arise from l.o.s. clouds between the Expanding Molecular Ring \citep[with a galactocentric radius of 200--300\,pc,][]{Whiteoak1979} and the 1\,kpc disc, while the lines with positive velocities arise from the envelope of Sgr\,B2(N) \citep{Greaves1994,Wirstrom2010,Thiel2019}. Specifically, our lines in the range $-$105 to $-$70\,km\,s$^{-1}$ are believed to be located within 1\,kpc of the GC \citep[][and references therein]{Corby2018}, and our lines with velocities around $+$60\,km\,s$^{-1}$ are associated with the envelope of Sgr\,B2(N) \citep{Wirstrom2010}.

This massive star-forming region harbours five hot cores, namely Sgr\,B2(N1--N5), with kinetic temperatures ranging from $\sim$130 to 150\,K for N3--N5, between 150 and 200\,K for N2, and between 160 and 200\,K for N1, assuming LTE conditions in all these cases \citep{Belloche2016,Bonfand2017,Belloche2019}. In addition, there are 20 $\sim$1.3\,mm continuum sources associated with dense clouds in Sgr\,B2(N) that exhibit a rich chemistry \citep{SanchezMonge2017,Schwoerer2019}. Recently, \citet{Ginsburg2018} detected 271 compact continuum sources at $\sim$3\,mm in the extended Sgr\,B2 cloud, thought to be high-mass protostellar cores, representing the largest cluster of high-mass young stellar objects reported to date in the Galaxy.

\begin{figure*}[h]
\includegraphics[width=1\linewidth, trim={0 110 0 0cm},clip]{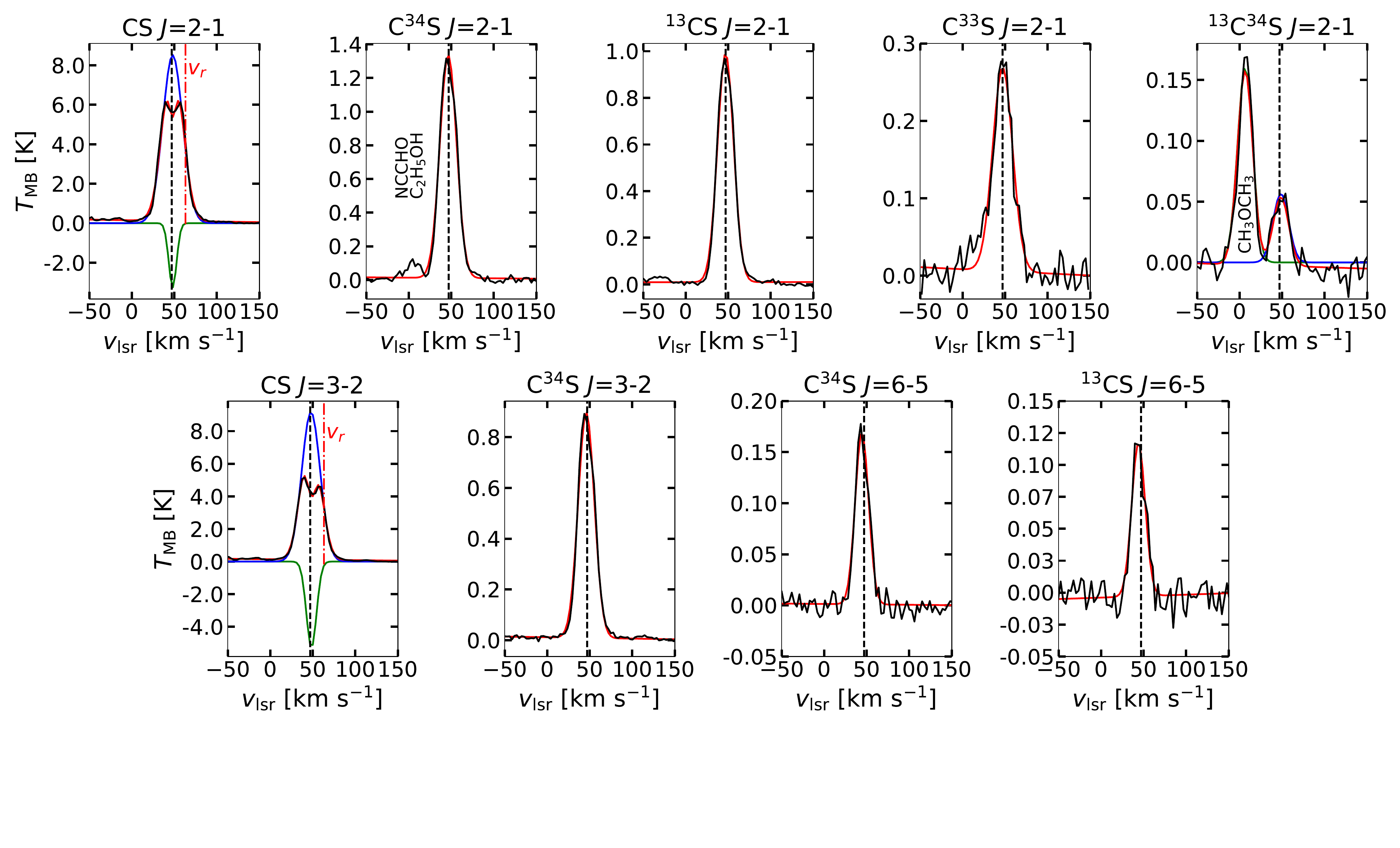}
\caption{From left to right and top to bottom: Observed line profiles of the \textit{J}=2--1 transitions of CS, C$^{34}$S, $^{13}$CS, C$^{33}$S, and $^{13}$C$^{34}$S, the \textit{J}=3--2 transitions of CS and C$^{34}$S, and the \textit{J}=6--5 transitions of C$^{34}$S and $^{13}$CS, in the $+$50\,km\,s$^{-1}$\,Cloud (EQ\,J2000: $17^\mathrm{h}45^\mathrm{m}50.20^\mathrm{s},-28^\mathrm{\circ} 59^\mathrm{\arcmin} 41.1^\mathrm{\arcsec}$). Measured profiles are shown in black; the resulting Gaussian fitting is presented in red. For the CS \textit{J}=2--1 and \textit{J}=3--2 lines, a double Gaussian fitting was performed to account for the self-absorbed component (in green) in an attempt to retrieve the undisturbed emission line (in blue) (see Sect.\,\ref{peak_opacities_and_line_intesity_ratios}). All the other lines are well fitted by a single component. In the case of $^{13}$C$^{34}$S \textit{J}=2--1, the companion line is subtracted because of its relatively strong emission and potential contamination. The dashed vertical lines denote the $^{13}$CS \textit{J}=2--1 peak position (46.94 km\,s$^{-1}$, see Table\,\ref{tab:2}). Dash-dotted red lines denote the redshifted velocity ($v_{r}$=63\,km\,s$^{-1}$) at which the CS \textit{J}=2--1 and \textit{J}=3--2 transition lines are not affected by self-absorption (see Sect.\,\ref{peak_opacities_and_line_intesity_ratios}). Nearby lines likely observed are labelled.}
\label{fig:50cloudimage}
\end{figure*}

\begin{figure*}
\centering

\includegraphics[,width=17cm, trim = 3.3cm 2.cm 2.35cm 12.5cm, clip=True]{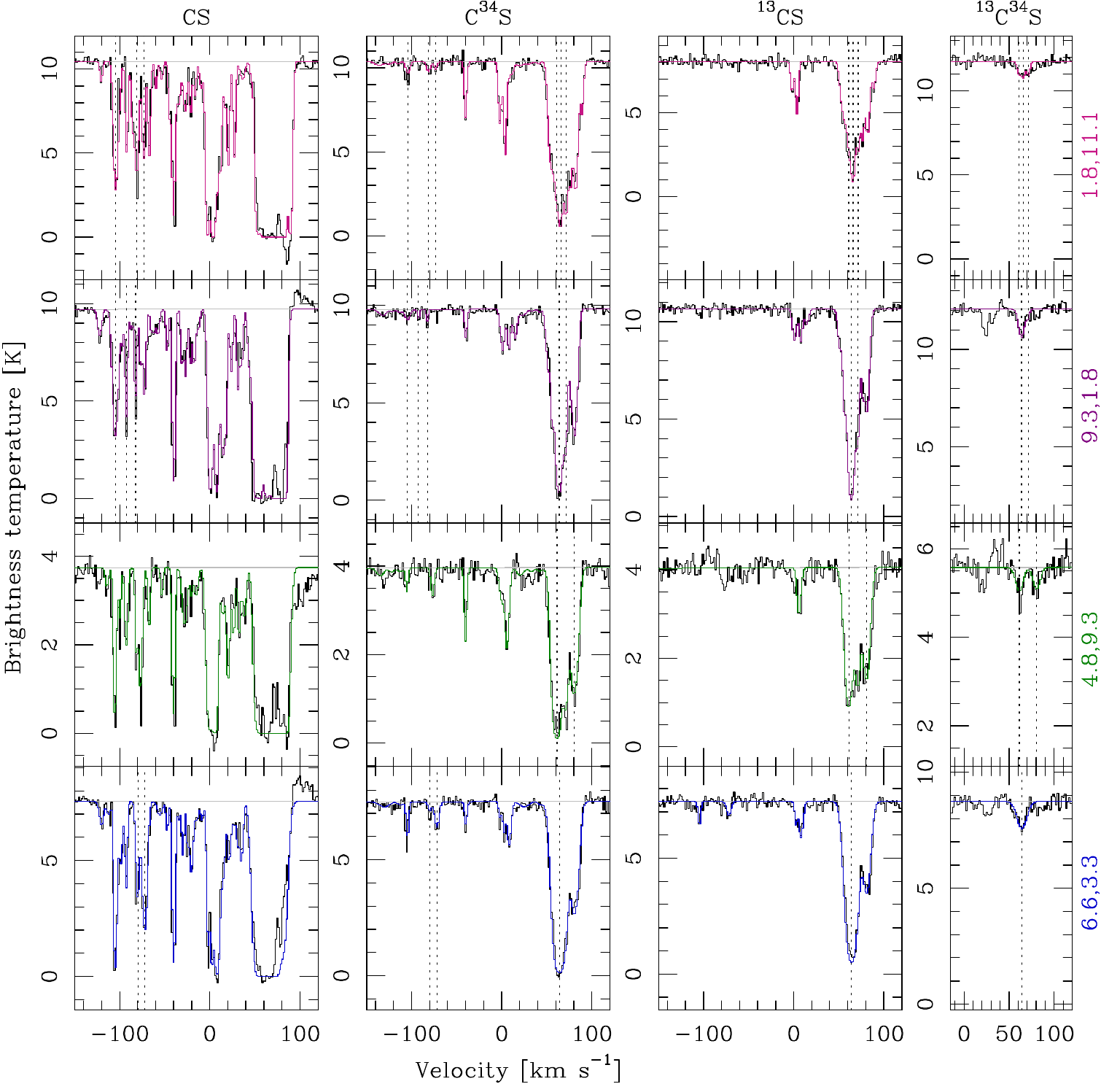}

\caption{EMoCA absorption spectra of four isotopologues of CS towards four positions in Sgr\,B2(N). The offsets (in units of arcseconds, see Table\,\ref{isotopic_ratios_emoca} and Fig.\,\ref{fig:SgrB2}) from the phase centre are indicated at the right of each row. The observed spectra are shown in black and the synthetic spectra in magenta, purple, green, and blue, depending on the position. The synthetic spectra were obtained by \citet{Thiel2019thesis}. The dotted vertical lines indicate the velocities at which we determined the isotopic ratios using the corresponding isotopologue. The velocities are listed in Table\,\ref{isotopic_ratios_emoca}. The grey horizontal lines show the
continuum level.}
\label{spectra_emoca}
\end{figure*}

\section{Results}
\label{results}

In the following we first discuss the measured profiles towards the $+$50\,km\,s$^{-1}$ Cloud, and provide the equations used to determine peak opacities and the carbon isotope ratio (Sect.\,\ref{peak_opacities_and_line_intesity_ratios}). Then we determine the $^{32}$S/$^{34}$S isotope ratio in two different ways (Sects.\,\ref{doubleisotopemethod} and \ref{ratiofromdirectobservations}) from the \textit{J}=2--1 lines of the different detected CS isotopologues. The \textit{J}=3--2 opacities and the $^{34}$S/$^{33}$S ratio are the topics of Sects.\,\ref{csandc34sopacities} and \ref{34S/33S}. Sgr\,B2(N) data are analysed in Sect.\,\ref{modellinSgrB2Ndata}, while the relation between the $^{32}$S/$^{34}$S isotope ratios from CS and the actual interstellar $^{32}$S/$^{34}$S values is the topic of Sect.\,\ref{fractionation}. Generally, the analysis assumes that lines with the same rotational quantum numbers, related to different CS isotopologues, are co-spatial. We present a summary of our results in Table\,\ref{tab:summary_50CLOUD}.

\subsection{\texorpdfstring{Peak opacities and column density ratios in the $+$50\,km\,s$^{-1}$ Cloud}{Peak opacities and column density ratios in the Cloud}}
\label{peak_opacities_and_line_intesity_ratios}
The CS emission lines observed towards the $+$50\,km\,s$^{-1}$\,Cloud are presented in Fig.\,\ref{fig:50cloudimage}. All lines show peaks in agreement with the local standard of rest (LSR) velocity of the system, $\sim$50\,km\,s$^{-1}$ \citep{Sandqvist2008,Requena-Torres2008}. In addition to the CS \textit{J}=2--1 isotopologue lines, there are probably weak features of cyanoformaldehyde (NCCHO) and ethanol (C$_{2}$H$_{5}$OH) at 96.4260958\,GHz and 96.4273380\,GHz, respectively, on the blue-shifted side ($\sim$0\,km\,s$^{-1}$) of C$^{34}$S; dimethyl ether (CH$_{3}$OCH$_{3}$) at 90.9375080\,GHz or $\sim$10\,km\,s$^{-1}$ on the blue-shifted side dominates the $^{13}$C$^{34}$S spectrum. Those molecules have been observed already in Sgr\,B2 \citep{Zuckerman1975,Nummelin1998,MartinPintado2001,Remijan2008,Belloche2013}. 

Moreover, the CS \textit{J}=2--1 and \textit{J}=3--2 line profiles show double-peaked profiles, which are readily explained by self-absorption, centred at the systemic velocity of the cloud in the \textit{J}=2--1 transition and with a self-absorption marginally redshifted (by $\sim$0.4\,km\,s$^{-1}$) in the \textit{J}=3--2 transition. The CS parameters allowed to vary freely and obtained from single- or double-component Gaussian fitting are summarised in Table\,\ref{tab:2}. They were obtained using a series of Python codes, mainly within the lmfit package\footnote{\url{https://lmfit.github.io/lmfit-py/intro.html}}. 

As can be seen in Fig.\,\ref{fig:50cloudimage}, for the CS \textit{J}=2--1 and \textit{J}=3--2 lines, we have fitted a double Gaussian considering a positive (in blue) and a negative (in green) Gaussian component each. Their parameters are summarised in Table\,\ref{tab:2}, where the uncertainties were taken directly from the lmfit package (for details see Appendix\,\ref{Gaussianfitting}).

Given that CS \textit{J}= 2--1 and \textit{J}= 3--2 show double-peaked profiles, while the rare isotopologues exhibit a single peak in between, the CS lines are likely optically thick. This was also suggested by \citet{Tsuboi1999} who find that CS \textit{J}= 1--0 is moderately optically thick in this object, with an opacity ($\tau$) of around 2.8. Then, we can further assume as a first approximation that C$^{34}$S \textit{J}= 2--1 is optically thin in the expected case that $^{32}$S/$^{34}$S\,$\gg$\,3 (see Sect.\,\ref{ratiofromdirectobservations} for a confirmation of this assumption; see also \citet{Frerking1980} and \citet{Corby2018}). In this case $^{13}$C$^{34}$S \textit{J}= 2--1 is definitively optically thin.

In the following we consider an excitation temperature range of 9.4--300\,K for our column density computations in order to obtain conservative estimates. Our column density values were calculated using Eq.\,80 in \citet{Mangum2015} assuming a filling factor of unity. We then deduce from the integrated line intensities of C$^{34}$S \textit{J}=2--1 and $^{13}$C$^{34}$S \textit{J}=2--1, converted to column densities, a $^{12}$C/$^{13}$C ratio of 22.1$^{+3.3}_{-2.4}$. This value is used in Equations \ref{eq:opacity} and \ref{eq:opacity2} (denoted R$_{\rm{C}}$), and it agrees with previous observations in the GC within the uncertainties \citep[$\sim$17-25; e.g.][]{Frerking1980,Corby2018}, and with the values derived from many transitions of complex organic molecules (COMs) in the hot core Sgr\,B2(N2) \citep{Belloche2016,Muller2016}, indicating that our approach is correct and confirming that C$^{34}$S is indeed optically thin.

Assuming equal excitation temperatures (see Appendix\,\ref{excitation_temperatures}) and beam filling factors for $^{12}$CS and $^{13}$CS, the $^{13}$CS \textit{J}=2--1 peak opacity $\tau$($^{13}$CS) can then be determined from
\begin{equation}
\label{eq:opacity}
    \frac{T_{\rm{MB}}(\rm{^{12}CS})}{T_{\rm{MB}}(\rm{^{13}CS})} = \frac{1 - e^{-\tau(\rm{^{13}CS})\rm{R_{C}}}}{1 - e^{-\tau(\rm{^{13}CS})}} , \rm{R_{C}} = \frac{\rm{^{12}C}}{\rm{^{13}C}}.
\end{equation}
However, as there is a self-absorption feature at the centre of the $^{12}$CS \textit{J}=2--1 and \textit{J}=3--2 profiles, we measure the line temperatures of both $^{12}$CS and $^{13}$CS at a redshifted velocity, $v_{r}$= 63 km\,s$^{-1}$, where the line shape is not affected by self-absorption. We can then retrieve the opacity at the systemic velocity, assumed to be the $^{13}$CS\,\textit{J}=2--1 peak velocity, $v_{sys}$=46.94\,km\,s$^{-1}$, by considering a Gaussian distribution. First, we compute the opacities at $v_{r}$ in the following way:

\begin{equation}
\label{eq:opacity2}
    \frac{T_{\rm{MB}}(\rm{^{12}CS_{\textit{v}_{\textit{r}}}})}{T_{\rm{MB}}(\rm{^{13}CS_{\textit{v}_{\textit{r}}}})} = \frac{1 - e^{-\tau(\rm{^{13}CS_{\textit{v}_{\textit{r}}}})\rm{R_{C}}}}{1 - e^{-\tau(\rm{^{13}CS_{\textit{v}_{\textit{r}}}})}} , \rm{R_{C}} = \frac{\rm{^{12}C}}{\rm{^{13}C}},
\end{equation}

\begin{equation}
\label{eq:develop}
\frac{4.26\pm 0.11}{0.24\pm0.11} = \frac{1 - e^{-22.1^{+3.3}_{-2.4}\tau(\rm{^{13}CS_{\textit{v}_{\textit{r}}}})}}{1 - \textit{e}^{-\tau(\rm{^{13}CS_{\textit{v}_{\textit{r}}}})}}.
\end{equation}

Here \textit{T}$_{\rm{MB}}(\rm{^{12}CS_{\textit{v}_{\textit{r}}}})$ and \textit{T}$_{\rm{MB}}(\rm{^{13}CS_{\textit{v}_{\textit{r}}}})$ are the main-beam brightness temperatures of CS and $^{13}$CS \textit{J}=2--1 at $v_{r}$. This results in $\tau(\rm{^{13}CS_{\textit{v}_{\textit{r}}}})$ = 0.02$\pm$0.01, considering the same uncertainties for the peak temperatures as those obtained by performing a single Gaussian fit in those lines. We can now retrieve the opacity at the systemic velocity as 

\begin{equation}
\label{eq:retrieve}
 \tau(\rm{^{13}CS_{\textit{v}_{\textit{sys}}}}) = \frac{\tau(\rm{^{13}CS_{\textit{v}_{\textit{r}}}})}{\textit{e}^{-(\textit{v}_{\textit{r}}-\textit{v}_{\textit{sys}})^2/2\sigma^2}}, 
\end{equation}

\noindent
where $\sigma$ is the full width at half maximum (FWHM) of $\rm{^{13}CS_{\textit{v}_{\textit{sys}}}}$ divided by $\sqrt{8ln(2)}$ (FHWM/$\sqrt{8ln(2)}$ = 9.46$\pm$0.12 km\,s$^{-1}$) and $v_{sys}$=46.942$\pm$0.118 km\,s$^{-1}$. We obtain $\tau(\rm{^{13}CS_{\textit{v}_{\textit{sys}}}})$ $=$ 0.08$^{+0.05}_{-0.04}$. Multiplying this by R$_{\rm{C}}$, we obtain $\tau(\rm{^{13}CS_{\textit{v}_{\textit{sys}}}})\rm{R}_{C}$= $\tau(\rm{CS_{\textit{v}_{\textit{sys}}}})$=1.9$^{+1.1}_{-0.8}$, consistent with previous observations \citep[$\tau(\rm{CS_{\textit{v}_{\textit{sys}}}})\sim$2.8;][]{Tsuboi1999}. The uncertainty on $v_{sys}$ corresponds to a 0.15\% variation in the $\tau(\rm{CS_{\textit{v}_{\textit{sys}}}}$) value in the worst case. Therefore, it is ignored in the following. 

\subsubsection{\texorpdfstring{A $^{32}$S/$^{34}$S ratio obtained through the double isotope method}{A 32S/34S ratio obtained through the double isotope method}}
\label{doubleisotopemethod}

As we have seen, CS must be moderately optically thick. Then, the $^{32}$S/$^{34}$S isotope ratios cannot be determined from the observed $N$($^{12}$CS)/$N$(C$^{34}$S) ratio. Instead, we can use the column densities of $^{13}$CS and C$^{34}$S by realistically assuming that those lines are optically thin (see Sect.\,\ref{peak_opacities_and_line_intesity_ratios}). Therefore, we have derived the values for $^{32}$S/$^{34}$S making use of the carbon isotope ratio mentioned above, in the following way:

\begin{equation}
\label{eq:4}
\frac{\rm{^{32}S}}{\rm{^{34}S}} = \frac{\rm{^{12}C}}{\rm{^{13}C}} \frac{N(\rm{^{13}CS})}{N(\rm{C^{34}S})}.
\end{equation}
From Eq.\,\ref{eq:4} we obtain a $^{32}$S/$^{34}$S \textit{J}=2--1 ratio of 16.3$^{+3.0}_{-2.4}$. By using the $^{13}$CS and C$^{34}$S \textit{J}=6--5 transitions, we obtain a $^{32}$S/$^{34}$S \textit{J}=6--5 ratio of 15.8$^{+4.2}_{-3.4}$. This agreement within the uncertainties can be taken as another argument in favour of the low opacity of C$^{34}$S and the subsequent validity of our assumptions and calculations. If some of the C$^{34}$S lines in the rotational ladder were not optically thin, we would expect different $N$($^{13}$CS)/$N$(C$^{34}$S) ratios in the \textit{J}=2--1 and 6--5 transitions due to photon trapping leading to higher excitation temperatures in the more abundant isotopologue, which is not observed.

\subsubsection{\texorpdfstring{$^{32}$S/$^{34}$S ratio from direct observations}{S32/S34 ratio from direct observations}}
\label{ratiofromdirectobservations}

As we have measured the $^{13}$C$^{34}$S \textit{J}=2--1 transition, we can also obtain the $^{32}$S/$^{34}$S ratio directly from
\begin{equation}
\label{eq:Sratiodirect}
      \frac{\rm{^{32}S}}{\rm{^{34}S}} = \frac{N(\rm{^{13}CS})}{N(\rm{^{13}C^{34}S})}.
\end{equation}
Using this we obtain a $^{32}$S/$^{34}$S \textit{J}=2--1 ratio of 16.3$^{+2.1}_{-1.7}$, consistent with the ratio obtained through the double-isotope method in Eq.\,\ref{eq:4} and again indicating that our initial assumptions concerning line saturation were correct. In the following, we use the latter value for our analysis because it was determined in the most direct way. In order to estimate the opacity of CS from C$^{34}$S (see Sect.\,\ref{csandc34sopacities}), we use this C$^{32}$S/C$^{34}$S
\textit{J}=2--1 ratio of 16.3$^{+2.1}_{-1.7}$ as the sulphur isotopic ratio $^{32}$S/$^{34}$S and call it R$_{\rm{S}}$.

To compare our results with those of \citet{Chin1996}, we also derived a sulphur ratio from the integrated intensities: $\rm{^{32}S}/\rm{^{34}S} \sim \textit{I}(\rm{^{13}CS})/\textit{I}(\rm{^{13}C^{34}S})$. This results in a $^{32}$S/$^{34}$S value of 18.6$^{+2.2}_{-1.8}$. The differences between the column density and integrated intensity ratios are due to the rotational partition functions, rotational constants, and Einstein A‐coefficients for spontaneous emission of radiation that slightly differ for the different isotopologues. 

\subsubsection{\texorpdfstring{CS and C$^{34}$S \textit{J}=3--2 opacities}{CS and C34S J=3--2 opacities}}
\label{csandc34sopacities}

Now we are able to determine the opacities of CS and C$^{34}$S \textit{J}=3--2 by proceeding in the same way as in Equations \ref{eq:opacity2} and \ref{eq:retrieve}, but considering this time the sulphur ratio. Here, as in Equation\,\ref{eq:opacity2}, we also assume equal excitation temperatures (see Appendix\,\ref{excitation_temperatures}) and beam filling factors, this time for $^{12}$CS and C$^{34}$S \textit{J}=3--2:
\begin{equation}
\label{eq:opacitywithsulfur}
    \frac{T_{\rm{MB}}(\rm{^{12}CS_{\textit{v}_{\textit{r}}}})}{T_{\rm{MB}}(\rm{C^{34}S_{\textit{v}_{\textit{r}}}})} = \frac{1 - e^{-\tau(\rm{C^{34}S_{\textit{v}_{\textit{r}}}})\rm{R_{S}}}}{1 - e^{-\tau(\rm{C^{34}S_{\textit{v}_{\textit{r}}}})}} , \rm{R_{S}} = \frac{\rm{^{32}S}}{\rm{^{34}S}}.
\end{equation}
From Eq.\,\ref{eq:opacitywithsulfur}, $\tau(\rm{C^{34}S_{\textit{v}_{\textit{r}}}})$=0.01--0.03. Following the formalism in Eq.\,\ref{eq:retrieve}, we obtain a $\tau(\rm{C^{34}S_{\textit{v}_{\textit{sys}}}})$ value of 0.05--0.15. Entering this value and replacing $\tau(\rm{C^{34}S_{\textit{v}_{\textit{sys}}}})$\rm{R$_{S}$} by $\tau(\rm{CS_{\textit{v}_{\textit{sys}}}})$, the CS$_{\textit{v}_{\textit{sys}}}$ opacity results in a proper range of 1.0--2.8, considering throughout this calculation that this latter value cannot lie below unity, since CS shows clear signs of optical thickness (see Sect\,\ref{peak_opacities_and_line_intesity_ratios}). This is consistent with previous observations \citep{Tsuboi1999}. All the derived opacities are summarised in Table\,\ref{tab:2}.

\subsubsection{\texorpdfstring{$^{34}$S/$^{33}$S ratio from direct observations}{S34/S33 ratio from direct observations}}
\label{34S/33S}

The C$^{33}$S \textit{J}=2--1 line was also observed. This offers the possibility of obtaining the $^{34}$S/$^{33}$S ratio for the GC. Since C$^{33}$S is less abundant than C$^{34}$S, we can expect a clearly optically thin profile. This ratio is easily obtained by
\begin{equation}
\label{eq:C33S}
    \frac{\rm{^{34}S}}{\rm{^{33}S}} = \frac{N(\rm{C^{34}S})}{N(\rm{C^{33}S})}.
\end{equation}
We obtain a $^{34}$S/$^{33}$S ratio of 4.3$\pm$0.2 for the $+$50\,km\,s$^{-1}$\,Cloud. If we take the integrated intensities instead of the column densities, this ratio would be 4.2$\pm$0.2, consistent with the lower end of the range of ratios obtained by \citet{Chin1996}, who derived $^{34}$S/$^{33}$S ratios between 4.38 and 7.53, irrespective of Galactic radius. Whether this is a first hint of a gradient remains to be seen. Better data from the Galactic disc are necessary to tackle this question.

\begin{table}[!h]
\caption{Summary for our carbon and sulphur column density ratio calculations in the $+$50\,km\,s$^{-1}$\,Cloud.}
\centering
\begin{tabular}{c c c c}
\hline
\multicolumn{1}{c}{$^{12}$C/$^{13}$C\tablefootmark{a}} & \multicolumn{1}{c}{$^{32}$S/$^{34}$S\tablefootmark{b}} & \multicolumn{1}{c}{$^{32}$S/$^{34}$S\tablefootmark{c}} & \multicolumn{1}{c}{$^{34}$S/$^{33}$S\tablefootmark{d}}\\

\multicolumn{1}{c}{\textit{J}=2--1} & \multicolumn{1}{c}{\textit{J}=2--1/\textit{J}=6--5} & \multicolumn{1}{c}{\textit{J}=2--1} & \multicolumn{1}{c}{\textit{J}=2--1} \\
\hline
\hline \\
22.1$^{+3.3}_{-2.4}$ & 16.3$^{+3.0}_{-2.4}$/15.8$^{+4.2}_{-3.4}$ & 16.3$^{+2.1}_{-1.7}$ & 4.3$\pm$0.2\\
\hline
\end{tabular}
\tablefoot{\tablefoottext{a}{From \textit{N}(C$^{34}$S)/\textit{N}($^{13}$C$^{34}$S) \textit{J}=2--1 lines, as described in Sect.\,\ref{peak_opacities_and_line_intesity_ratios}.\tablefoottext{b}{Through the double isotope method in Sect.\,\ref{doubleisotopemethod}.}\tablefoottext{c}{From direct observations in Sect.\,\ref{ratiofromdirectobservations}.}\tablefoottext{d}{Sect.\,\ref{34S/33S}.}}}
\label{tab:summary_50CLOUD}
\end{table}

\subsection{Modelling the Sgr B2(N) data} 
\label{modellinSgrB2Ndata}

\begin{table*}[!h]
\caption{Isotopic ratios determined in the envelope of Sgr\,B2(N) ($^{13}$CS/$^{13}$C$^{34}$S and C$^{34}$S/$^{13}$C$^{34}$S) and in GC clouds along the line of sight to Sgr\,B2(N) (CS/C$^{34}$S) using absorption lines of CS isotopologues.}
\label{isotopic_ratios_emoca}
\subcaption*{Isotopic ratios determined in the envelope of Sgr\,B2(N)}
\centering
\setlength{\tabcolsep}{2.5pt}
\begin{tabular}{c c r r r r r r r}       
\hline               
$\Delta x$\tablefootmark{a} & $\Delta y$\tablefootmark{a} & \multicolumn{1}{c}{\textit{V}$_{LSR}$} & \multicolumn{1}{c}{\textit{N}(C$^{34}$S)\tablefootmark{b}} & \multicolumn{1}{c}{\textit{N}($^{13}$CS)\tablefootmark{b}} & \multicolumn{1}{c}{\textit{N}($^{13}$C$^{34}$S)\tablefootmark{b}} & \multicolumn{1}{c}{FWHM\tablefootmark{c}} & \multicolumn{1}{c}{$^{13}$CS/$^{13}$C$^{34}$S} & \multicolumn{1}{c}{C$^{34}$S/$^{13}$C$^{34}$S}\\

[\arcsec] & [\arcsec] & [km\,s$^{-1}$] & [10$^{12}$\,cm$^{-2}$] & [10$^{12}$\,cm$^{-2}$] & [10$^{12}$\,cm$^{-2}$] & \multicolumn{1}{c}{[km\,s$^{-1}$]} & & \\
\hline
\hline  \\
1.8	&	11.1	&	71.6	&	$70.0\pm1.1$      &   $35.0\pm1.2$    &    $2.5\pm0.6$	&	4.0		&$14.0\pm3.3$	&	$28.0\pm6.6$	\\
1.8	&	11.1	&	65.8	&	$120.0\pm1.6$     &   $90.0\pm2.2$    &    $3.5\pm0.7$  &	5.0		&$25.7\pm5.3$	&	$34.3\pm7.1$	\\
1.8	&	11.1	&	61.1	&	$65.0\pm1.1$      &   $38.0\pm1.4$    &    $2.6\pm0.7$  &	5.0		&$14.6\pm4.1$	&	$25.0\pm7.0$	\\

	&	    &       &		&		&		&		&		&		\\
9.3	&	1.8	&	71.1	&	$39.0\pm0.9$    &   $30.0\pm0.9$    &    $1.2\pm0.4$&	4.0 &	$25.0\pm8.8$	&	 $32.5\pm11.4$	\\
9.3	&	1.8	&	63.8	&	$200.0\pm2.3$   &  $160.0\pm2.1$    &    $8.2\pm0.8$   &	7.5	&	$19.5\pm2.0$	&	$24.4\pm2.4$	\\

	&		&		&		&		&		&		&		&   	\\
4.8	&	9.3	&	80.6	&	$94.0\pm3.1$    &   $74.0\pm3.7$    &    $7.0\pm1.9$  &	9.0	&	$10.6\pm3.0$	&	$13.4\pm3.7$	\\
4.8	&	9.3	&	61.3	&	$280.0\pm6.5$   &  $120.0\pm4.7$    &    $8.0\pm2.0$  &	9.5	&	$15.0\pm3.8$	&	$35.0\pm8.8$	\\
	&		&		&		&		&		&		&		&		\\
6.6	&	3.3	&	64.2	&	$420.0\pm4.5$   &  $280.0\pm4.3$    &    $15.0\pm1.5$	&	12.5	&	$18.7\pm1.8$	&	$28.0\pm2.7$	\\

\multicolumn{2}{c}{average}    & & & &  &         &    $17.9\pm5.0$    &    $27.6\pm6.5$    \\
\hline                      
\end{tabular}

\bigskip

\subcaption*{Isotopic ratios determined in GC clouds along the line of sight to Sgr\,B2(N)}
\centering
\setlength{\tabcolsep}{11.7pt}
\begin{tabular}{c c r r r r r}       
\hline               
$\Delta x$\tablefootmark{a} & $\Delta y$\tablefootmark{a} & \multicolumn{1}{c}{\textit{V}$_{LSR}$} & \multicolumn{1}{c}{\textit{N}(CS)\tablefootmark{b}} & \multicolumn{1}{c}{\textit{N}(C$^{34}$S)\tablefootmark{b}} & \multicolumn{1}{c}{FWHM\tablefootmark{c}} & \multicolumn{1}{c}{CS/C$^{34}$S}\\

[\arcsec] & [\arcsec] & [km\,s$^{-1}$] & [10$^{12}$\,cm$^{-2}$] & [10$^{12}$\,cm$^{-2}$] & [km\,s$^{-1}$] & \\
\hline
\hline  \\

1.8	&	11.1	&	-73.3	&	$26.0\pm0.7$    &   $1.4\pm0.5$	&	3.5	&	$18.6\pm6.3$		\\
1.8	&	11.1	&	-81.2	&	$45.0\pm0.9$    &   $2.5\pm0.6$	&	5.0	&	$18.0\pm4.0$		\\
1.8	&	11.1	&	-104.2	&	$58.0\pm1.0$    &   $3.0\pm0.6$	&	5.0	&	$19.3\pm3.6$		\\
	&	    &           &		    &	    	&		&	    	\\
9.3	&	1.8	&	-82.5	&	$16.0\pm0.7$        &  $1.4\pm0.4$	&	2.5	&	$11.4\pm3.7$		\\
9.3	&	1.8	&	-92.6	&	$20.0\pm0.7$        &  $1.2\pm0.5$	&	2.5	&	$16.7\pm7.3$		\\
9.3	&	1.8	&	-104.7	&	$63.0\pm1.1$        &  $3.0\pm0.7$	&	6.5	&	$21.0\pm4.9$		\\
	&		&	    	&	    	&	    	&		&				\\
6.6	&	3.3	&	-71.8	&	$65.0\pm1.5$        &  $7.2\pm0.8$	&	5.5	&	$9.0\pm1.0$		    \\
6.6	&	3.3	&	-79.8	&	$29.0\pm1.0$        &  $1.8\pm0.6$	&	4.0	&	$16.1\pm5.8$		\\
\multicolumn{2}{c}{average}    & &  &  & &    $16.3\pm3.8$         \\
\hline                      
\end{tabular}
\tablefoot{\tablefoottext{a}{The offset positions ($\Delta x$, $\Delta y$) in units of arcseconds: (1.8, 11.1), (9.3, 1.8), (4.8,9.3), and (6.6, 3.3), correspond to K4, K6$_\mathrm{shell}$, K5$_\mathrm{shell}$, and K6$_\mathrm{shell, a}$, in Fig.\,\ref{fig:SgrB2}, respectively. See the caption to Fig.\,\ref{fig:SgrB2} and the green crosses in the image.}\tablefoottext{b}{Column densities determined using Weeds.}\tablefoottext{c}{It is
assumed that all isotopologues have the same \textrm{FWHM}}. The average isotope ratios presented by the lowest line of each panel are unweighted and provide the standard deviation of an individual measurement (without dividing by the square root of the number of ratios).}
\end{table*}

Here we rely on the modelling of the absorption profiles of the isotopologues of CS carried out by \citet{Thiel2019thesis} using the EMoCA survey, following the same method as \citet{Thiel2019}. They used the software Weeds \citep{Maret2011} to model the absorption profiles. Their work assumes that all transitions of a molecule have the same excitation temperature and that the beam filling factor is unity, which is a reasonable assumption given that most absorption features are extended on scales of 15\arcsec\,or beyond in the ALMA maps \citep[see][their Sect. 5.4]{Thiel2019}, while the beam size is 1\farcs6 (see Sect.\,\ref{los}, Fig.\,\ref{fig:SgrB2}, and Table\,\ref{tab:1}). The fitted parameters were the column density, line width, and the centroid velocity, under the assumption that the excitation temperature is equal to the temperature of the cosmic microwave background (2.73 K). 

We selected four continuum peaks inside Sgr\,B2(N) (Fig.\,\ref{fig:SgrB2}). We excluded the two strong continuum peaks at which the main hot cores N1 and N2 are located because at these positions the spectra are full of emission lines of organic molecules \citep[e.g.][]{Bonfand2017} contaminating the carbon monosulfide absorption features. The offsets to the centre of the observed field are (1\farcs8, 11\farcs1), (9\farcs3, 1\farcs8), (4\farcs8, 9\farcs3), and (6\farcs6, 3\farcs3) (see Fig.\,\ref{fig:SgrB2} and Table\,\ref{isotopic_ratios_emoca}). The observed absorption profiles and the corresponding Weeds models for the four isotopologues and the four positions are shown in Fig.\,\ref{spectra_emoca}. Using their results for the column densities, we determined the isotopic ratios CS/C$^{34}$S, $^{13}$CS/$^{13}$C$^{34}$S, and C$^{34}$S/$^{13}$C$^{34}$S. We determined those ratios separately for the envelope of Sgr\,B2(N) and some GC clouds along the l.o.s. to Sgr\,B2(N), the latter with velocities lower than $-50$\,km\,s$^{-1}$. We only determine the ratio CS/C$^{34}$S in those cases where the absorption caused by CS is not optically thick.

The resulting unweighted average values of the isotopic ratios are listed in Table\,\ref{isotopic_ratios_emoca}, namely a $^{32}$S/$^{34}$S isotope ratio of 16.3$\pm$3.8 in the GC l.o.s. clouds towards Sgr\,B2(N) and 17.9$\pm$5.0 in the envelope of Sgr\,B2(N). For this envelope we obtain a $^{12}$C/$^{13}$C ratio of 27.6$\pm$6.5. It should be noted that our uncertainties correspond to the standard deviation for independent measurements, i.e. without dividing it by the square root of the number of studied spectral components.

\subsection{\texorpdfstring{Discussion on the validity of using C$^{32}$S/C$^{34}$S as a proxy for $^{32}$S/$^{34}$S}{Discussion on the validity of using C32S/C34S as a proxy for S32/S34}}
\label{fractionation}

\citet{Chin1996} estimated that sulphur fractionation is marginal for CS isotopic ratios. If the bulk of the CS emission, which allows us to measure rare isotopes, arises from the densest parts of the molecular clouds only, the heating from the massive stars should inhibit significant fractionation \citep{Chin1996}. In that case, CS emission can be used directly to determine sulphur isotope ratios from such sources.

\begin{figure*}[hbt!]
\begin{minipage}{.33\linewidth}
\centering
\includegraphics[width=0.96\linewidth, trim = 1.cm 1.6cm 0.cm 0.cm, clip=True]{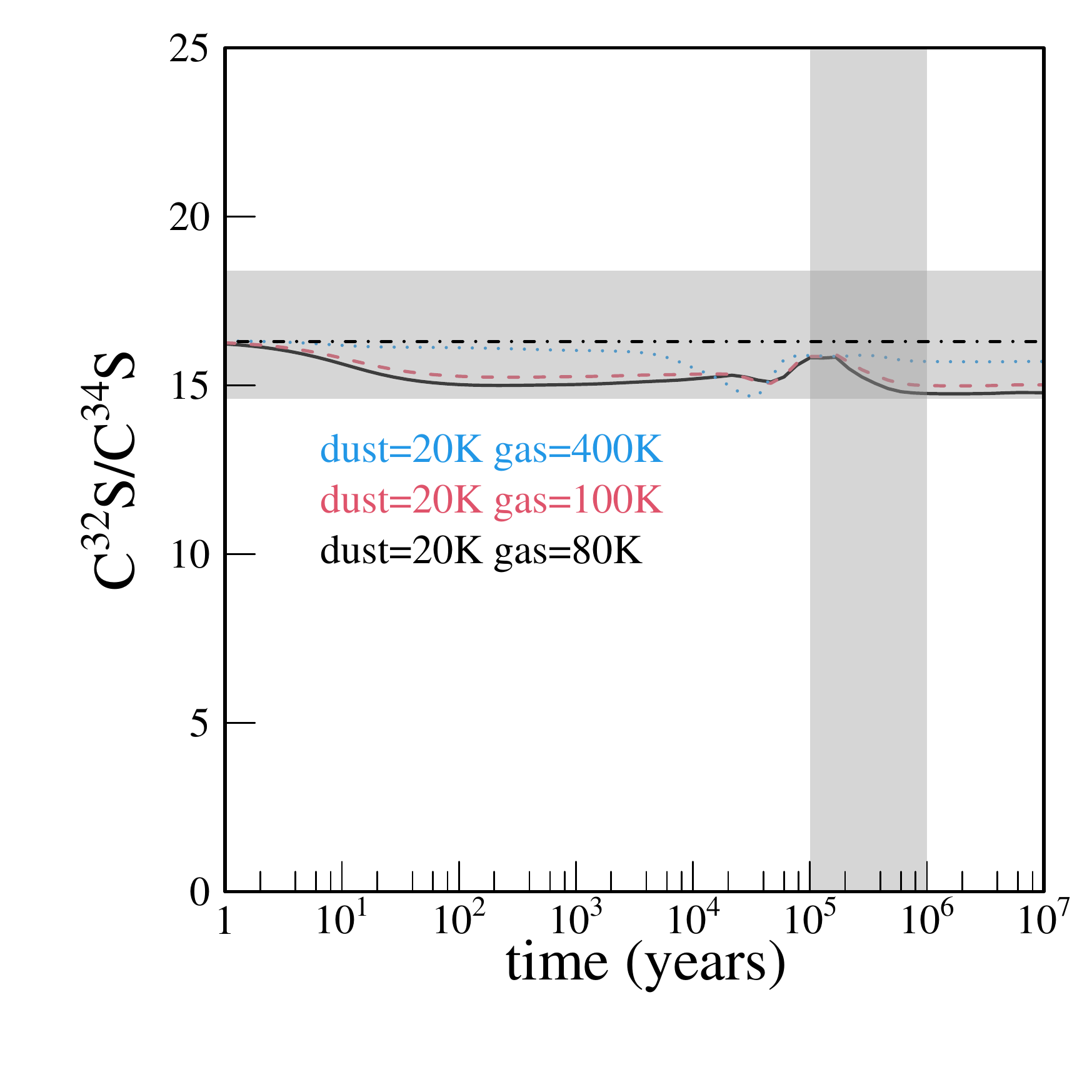}
\subcaption{} 
\end{minipage}
\begin{minipage}{.33\linewidth}
\centering
\includegraphics[width=0.96\linewidth, trim = 1.cm 1.6cm 0.cm 0.cm, clip=True]{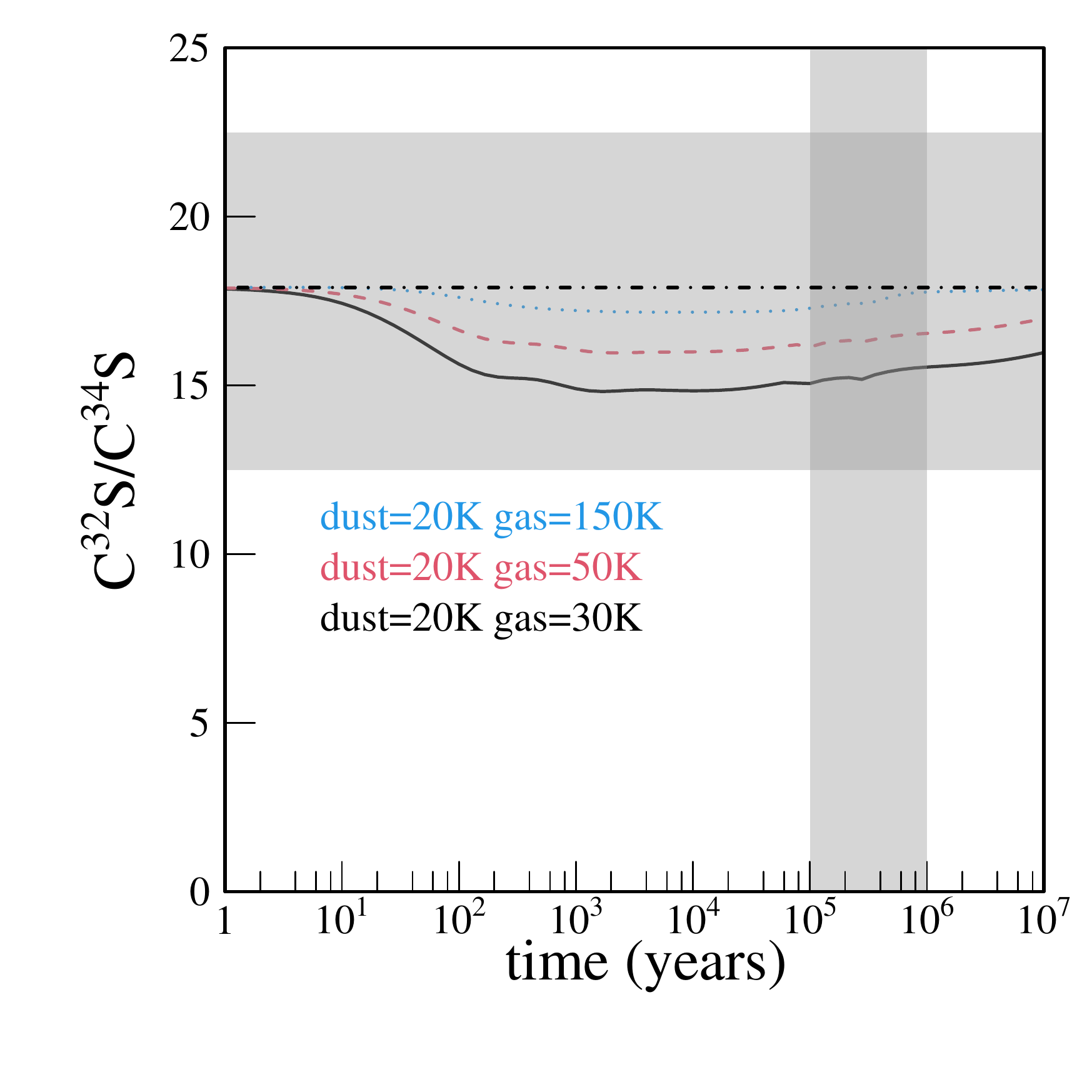}
\subcaption{} 
\end{minipage}%
\begin{minipage}{.33\linewidth}
\centering
\includegraphics[width=0.96\linewidth, trim = 1.cm 1.6cm 0.cm 0.cm, clip=True]{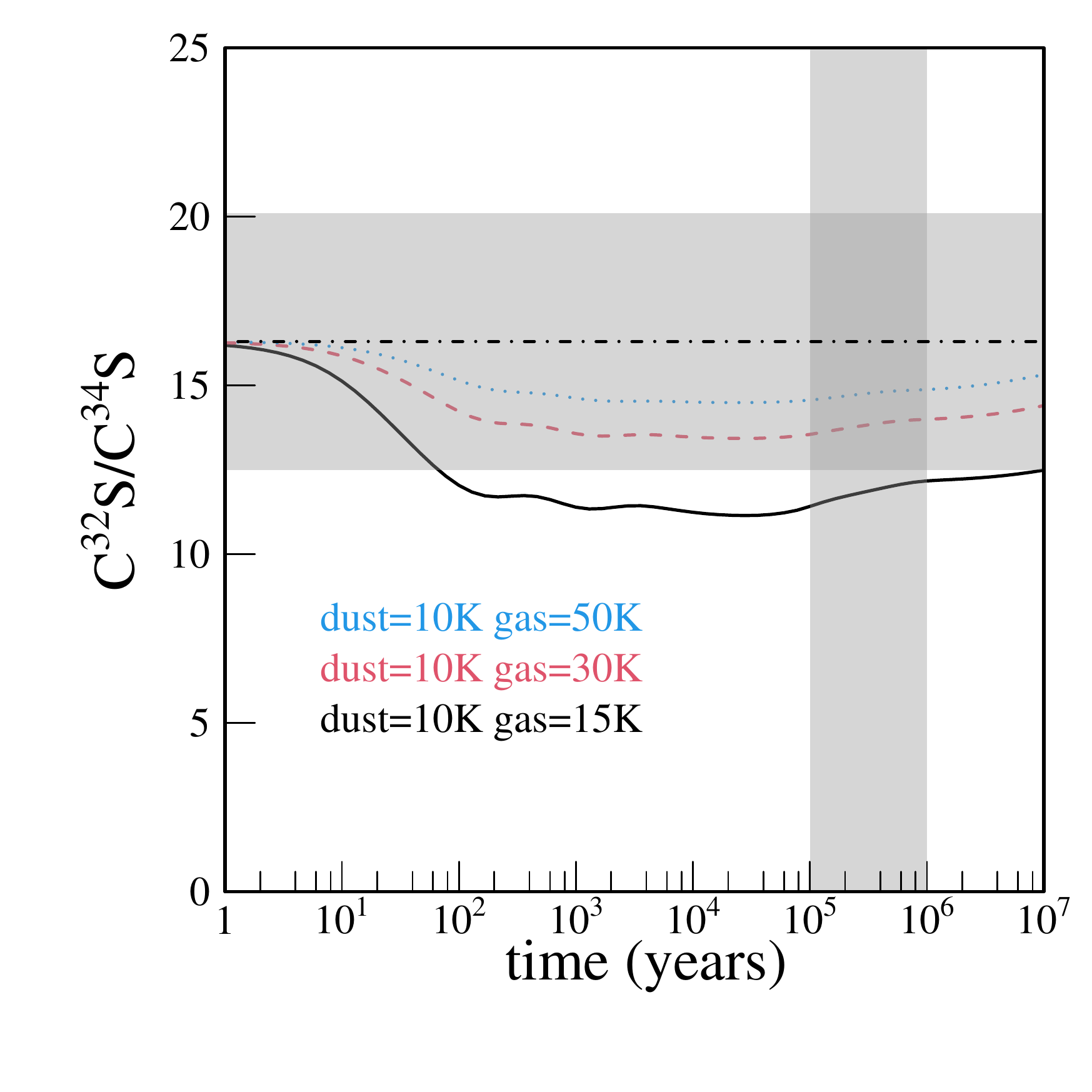}
\subcaption{} 
\end{minipage}%
\caption{Calculated abundance ratios of gas phase species C$^{32}$S/C$^{34}$S as a function of cloud age for conditions in (a) the $+$50\,km\,s$^{-1}$\,Cloud, (b) the envelope of Sgr\,B2(N), and (c) l.o.s. clouds towards Sgr\,B2(N) (see Sect.\,\ref{modellinSgrB2Ndata} for details). Low values for the gas temperatures were chosen to illustrate an upper limit for fractionation. The vertical grey loci represent values given by the most probable chemical age. The observational results from this study are illustrated as horizontal light grey rectangles (including the uncertainties).}
\label{models}
\end{figure*}

In their oxygen fractionation study, \citet{Loison2019} analyse sulphur fractionation including CS. Some sulphur fractionation is induced at low temperature by the $^{34}$S$^{+}$ $+$ CS $\rightarrow$ S$^{+}$ $+$ C$^{34}$S reaction. To determine the potential fractionation of sulphur, we used the network from \citet{Loison2019} in the $+$50\,km\,s$^{-1}$\,Cloud, the l.o.s. clouds towards Sgr\,B2(N) and the envelope of Sgr\,B2(N), with realistic physical conditions for these objects, in particular a much higher value of the cosmic-ray ionisation rate (CRIR) than the usual value in more local dense molecular clouds. Some typical results are shown in Fig.\,\ref{models} and are described below. 

In the simulations all elements with an ionisation potential below the maximum energy of ambient UV photons (13.6\,eV) are assumed to be initially in an atomic, singly ionised state. We considered all sulphur in the S$^{+}$ form without depletion, and we performed some tests to quantify the effect of depletion, which is low (see below). Hydrogen, with its high degree of self-shielding, is taken to be entirely molecular. The initial abundances are similar to those in Table\,1 of \citet{Hincelin2011}, the C/O elemental ratio being equal to 0.7 in our study. We verified that the initial state of carbon and nitrogen (C$^{+}$ versus CO and N versus N$_2$) have very little influence on sulphur fractionation (less than 4\% for the typical ages considered: 10$^{5-6}$ years). The estimation of the dense cloud ages is deduced from clouds with similar density (10$^{4}$ to a few 10$^{5}$\,cm$^{-3}$) for which the age is given by the best agreement between calculations and observations for key species given by the so-called distance of disagreement \citep{Wakelam2006}. By key species we mean species typically encountered in molecular clouds such as HCN, HNC, CN, CH, C2H, $c$--C3H, $c$--C3H2, CO, H$_{2}$CO, CH$_{3}$OH, NO, SO, CS, HCS$^{+}$, and H$_{2}$CS \citep[see e.g. ][]{Wakelam2010, Agundez2013,Agundez2019}.

For Fig.\,\ref{models}a, which represents conditions in the $+$50\,km\,s$^{-1}$\,Cloud, we adopted a density of 10$^{5}$\,cm$^{-3}$ and a CRIR of $\zeta ^{\rm{H_{2}}}$ = 7$\times$10$^{-16}$\,s$^{-1}$ based on measurements in hot cores of Sgr\,B2(N) by using COMs \citep{Bonfand2019}. For Fig.\,\ref{models}b and c, which represent the conditions for the envelope of Sgr\,B2(N) and the l.o.s. clouds towards Sgr\,B2(N), respectively, we chose a density of 10$^{4}$\,cm$^{-3}$, an upper limit for the volume density in those regions \citep[see][their Table 12]{Thiel2019}, in order to avoid possible UV heating in our models. Due to this high density, we have adopted a CRIR of $\zeta ^{\rm{H_{2}}}$ = 3$\times$10$^{-15}$\,s$^{-1}$, i.e. one order of magnitude lower than the value usually obtained in the l.o.s. of translucent and diffuse clouds towards the Galactic centre, but within the range obtained for the nuclear $\sim$100\,pc of our Galaxy \citep{Indriolo2015,LePetit2016}. The $^{32}$S/$^{34}$S isotope ratio chosen for each simulation is that obtained from our measurements, namely 16.3$^{+2.1}_{-1.7}$, 17.9$\pm$5.0, and 16.3$\pm$3.8 for Figs.\,\ref{models}a, b, and c, respectively.

Despite the limited literature on the subject, according to models from \citet{Laas2019}, the $+$50\,km\,s$^{-1}$\,Cloud is the only object in this study that could show signs of sulphur depletion. However, for a depletion level of up to 90\%, our models give results consistent with no sulphur depletion (which is the case for the results shown in Fig.\,\ref{models}) because of the high temperatures that prevent any efficient fractionation. It should be noted that if this cloud's chemistry was determined in an earlier colder evolutionary period with possibly significant sulphur fractionation, CS, in contrast to CO, could not accumulate because it would have been destroyed by protonation as the dissociative recombination of HCS$^{+}$ leads mainly to S $+$ CH and not to CS $+$ H (see Appendix\,\ref{chemicalaspects}). Therefore, the memory effect for the CS fractionation of such a dense cloud is small. For the l.o.s. clouds towards Sgr\,B2(N) and the envelope of Sgr\,B2(N), some runs (those shown in Fig.\,\ref{models}) give some sulphur fractionation when the gas temperature is low. This is due to the combination of low density limiting the depletion of sulphur and a high CRIR  ($\zeta ^{\rm{H_{2}}}$ = 3$\times$10$^{-15}$\,s$^{-1}$). These characteristics induce an elevated concentration of S$^+$ in the gas phase and then a $^{34}$S enrichment through the $^{34}$S$^{+}$ $+$ CS $\rightarrow$ S$^{+}$ $+$ C$^{34}$S reaction \citep[see Table 2 in][]{Loison2019}. It should be noted that the cases with some $^{34}$S enrichment only concern low kinetic temperatures, and that this relatively low enrichment will be even lower with some sulphur depletion. This depletion is not well constrained in the $+$50\,km\,s$^{-1}$\,Cloud, the l.o.s. clouds towards Sgr\,B2(N), and the envelope of Sgr\,B2(N), but it is high for some cold dense molecular clouds (between 10 and 25 for the dark cloud L1544 Barnard 1b \citep{Fuente2016} and even up to 200 for the dark cloud L1544 \citep{Vastel2018}.

The extremely low gas temperature cases in our models (black lines in Fig.\,\ref{models}) are shown to demonstrate the dependence of sulphur depletion on gas temperature. Since Galactic centre molecular cloud temperatures are higher \citep{Ginsburg2016}, CS shows likely very little fractionation in $^{34}$S for the purposes of this study and our measurements of the C$^{32}$S/C$^{34}$S ratio are a good approximation to the $^{32}$S/$^{34}$S ratio (even if the values thus obtained may be slightly underestimated when considering the results of the models). We can then expect a $\lesssim$10\% increase in the $^{32}$S/$^{34}$S ratios obtained for the $+$50\,km\,s$^{-1}$\,Cloud in Fig.\,\ref{models}a and slightly more for the l.o.s. clouds towards Sgr\,B(N) and its envelope ($\lesssim$15\%), considering the conditions labelled in red and blue in Figs.\,\ref{models}b and c.

Some measurements using the double-isotope ratio method (see Sect.\,\ref{doubleisotopemethod}), which take the $^{12}$C/$^{13}$C ratio into account, may induce a bias since CS may show a non-negligible fractionation into $^{13}$C. There is no specific study of the $^{13}$C fractionation of CS, but the reactivity of C$^{+}$ and C, in particular with CO and CN (but also with CS), can induce an enrichment or depletion in $^{13}$C of carbonaceous species including CS \citep{Smith1980,Roueff2015}. In that case, the good agreement between the sulphur fractionation measurements using C$^{32}$S/C$^{34}$S and the values obtained using the double-isotope method also suggests a low $^{13}$C fractionation of CS. This result is interesting and could initiate future studies on the modelling of $^{13}$C fractionation in CS.

\section{Our results in the light of previous studies}
\label{previous studies}

If we assume that the gradient proposed by \citet{Chin1996} would also be valid in the Galactic centre region, the $^{32}$S/$^{34}$S ratio would decrease to values of 4.1$\pm$3.1 at the centre of our Galaxy, i.e. to a very low value, only 1/4 of the solar system ratio. This value is less than one fourth of the value derived from integrated intensities in this work, 18.6$^{+2.2}_{-1.8}$. 

This difference can be explained in terms of the $^{12}$C/$^{13}$C ratios assumed in \citet{Chin1996}, required to obtain the $^{32}$S/$^{34}$S ratio through the double-isotope method by using the formalism of Eq.\,\ref{eq:4} (although using intensities instead of column densities). The $^{12}$C/$^{13}$C ratios were derived from the relation found by \citet{Wilson1994} of $^{12}\rm{C}/^{13}\rm{C} = (7.5 \pm 1.9)(\textit{D}\textsubscript{GC}/\rm{kpc}) + (7.6 \pm 12.9)$ that gives a value of 7.6$^{+12.9}_{-7.6}$ for the Galactic centre, although the authors claimed a value of $\sim$20 near the Galactic nucleus \citep[Sect 5.1]{Wilson1994}. This provides an idea of the large uncertainty in this relation. 

On the other hand, we are confident about our $^{12}$C/$^{13}$C ratio of 22.1$^{+3.3}_{-2.4}$ (Sect.\,\ref{peak_opacities_and_line_intesity_ratios}) for two reasons: first, the agreement between the $^{32}$S/$^{34}$S ratio obtained through the double-isotope method (Sect.\,\ref{doubleisotopemethod}), which makes use of the $^{12}$C/$^{13}$C ratio, and the $^{32}$S/$^{34}$S ratio obtained directly from $^{13}$CS/$^{13}$C$^{34}$S (Sect.\,\ref{ratiofromdirectobservations}), i.e. independently of the carbon ratio (this also indicates that the carbon fractionation is low, as described in Sect.\,\ref{fractionation}) and second, its proximity to the ratio obtained through decades of observations in the nuclear regions of our Galaxy \citep[$^{12}$C/$^{13}$C = 17--25,][]{Frerking1980,Wilson1994,Milam2005,Muller2008,Corby2018}, including LTE modelling of complex organic molecules \citep{Belloche2016,Muller2016}. There is no abrupt redirection in the $^{12}$C/$^{13}$C ratio \citep[e.g.][]{Henkel1985}. 

Recently, \citet{Corby2018} found $^{32}$S/$^{34}$S ratios mostly in the 5--10 range, based on C$^{32}$S and C$^{34}$S \textit{J}=1--0 absorption lines from diffuse clouds near the GC, with a resolution of $\sim$15\arcsec. Considering their data in the --73 to --106 km\,s$^{-1}$ velocity range, corresponding to our GC l.o.s. clouds towards Sgr\,B2(N), their observations reach values between 6.6$\pm$6 and 29$\pm$14, consistent with our values between 9.0$\pm$1.0 and 21.0$\pm$4.9 (Table\,\ref{isotopic_ratios_emoca}, lower panel).

In addition, \citet{ArmijosAbendano2015}, with a resolution of $\sim$30\arcsec--38\arcsec\,, found values of $\gtrsim$22 and 8.7$\pm$1.3 for $^{32}$S/$^{34}$S isotope ratios in l.o.s. clouds towards Sgr\,A and Sgr\,B2, respectively, consistent with previous estimations \citep{Frerking1980}. However, their sulphur ratios were obtained from OCS/OC$^{34}$S, with OCS being potentially optically thick and OC$^{34}$S spectra being badly affected by band pass ripples, possibly providing only tentative detections. So we propose a more conservative lower limit for the l.o.s. clouds towards Sgr\,A of $\sim$10 and we suggest that the uncertainty for their ratio in Sgr\,B2 was underestimated. 

Both \citet{Corby2018} and \citet{ArmijosAbendano2015} employed integrated column density ratios, so we should compare those measurements with our normal $^{32}$S/$^{34}$S isotope ratio estimation, that is 16.3$^{+2.1}_{-1.7}$ for the $+$50\,km\,s$^{-1}$ Cloud (as an approximation for their l.o.s. clouds towards Sgr A) and 16.3$\pm$3.8 for the l.o.s. clouds towards Sgr\,B2 (see Table\,\ref{isotopic_ratios_emoca}). Our data represent a significant improvement in terms of accuracy and precision with respect to those previous observations. In addition, our estimation of 17.9$\pm$5.0 for the envelope of Sgr\,B2(N) agrees with both estimations for the $+$50\,km\,s$^{-1}$ Cloud and also with previous calculations for the whole Sgr\,B2 region: $\sim$16, from the OCS/OC$^{34}$S ratio, which is claimed to be derived from optically thin lines \citep{Goldsmith1981}.

Cutting-edge model calculations performed by \citet{Kobayashi2011} relate sulphur isotope ratios ($^{32}$S/$^{33,34,36}$S) with metallicity ([Fe/H]\footnote{where [Fe/H] = log$_{10}$([$N_{\rm{Fe}}$/$N_{\rm{H}}$])$_{\rm{star}}$ - log$_{10}$([$N_{\rm{Fe}}$/$N_{\rm{H}}$])$_{\rm{sun}}$}). We can use this relation in combination with a given metallicity gradient along the Milky Way ([Fe/H] versus \textit{D}\textsubscript{GC}/kpc), to derive a $^{32}$S/$^{34}$S versus \textit{D}\textsubscript{GC}/kpc relation and compare it with our measurements.

\begin{figure*}
  \includegraphics[width=\linewidth]{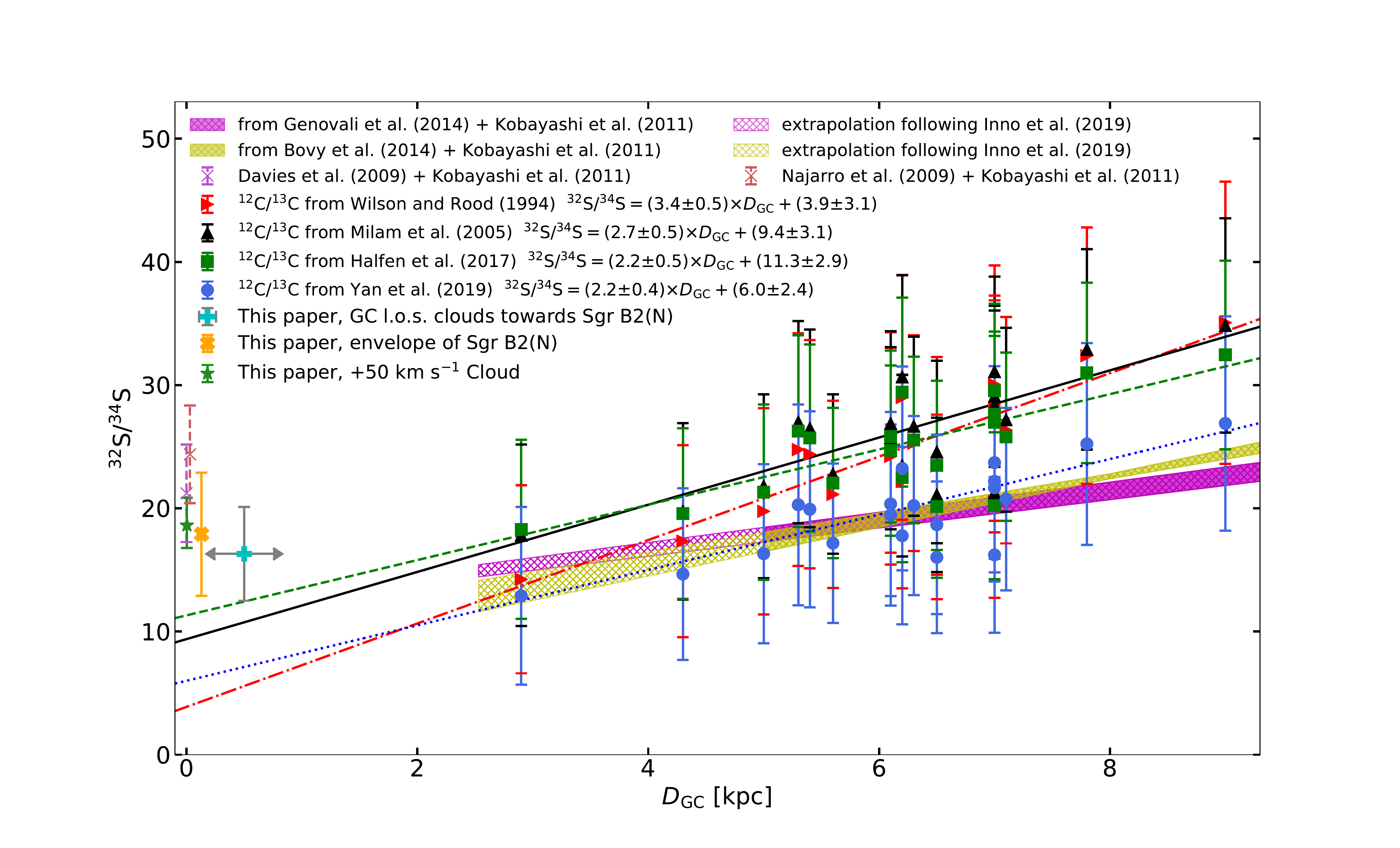}
  \caption{Sulphur isotope $^{32}$S/$^{34}$S ratio variation when accounting for different carbon $^{12}$C/$^{13}$C ratios as a function of galactocentric radius, $D$\textsubscript{GC} \citep[e.g.][]{Wilson1994, Halfen2017, Milam2005, Yan2019} \citep[for the implemented $^{13}$C$^{32}$S/C$^{34}$S ratios, see][]{Chin1996}. The $^{32}$S/$^{34}$S to $D$\textsubscript{GC} relations obtained from a linear least-squares fit to weighted data (taken as 1/$\sigma$, see Sect.\,\ref{previous studies}) are shown and plotted as lines with different styles (see legend). The $^{32}$S/$^{34}$S ratios from this work are shown in orange, cyan, and light green. All ratios were gleaned from integrated intensity ratios except for Sgr\,B2(N) (see Fig.\,\ref{spectra_emoca}), where our derived integrated column density ratios are used. Possible differences between integrated column density ratios and line intensity ratios for the mean values in Sgr\,B2(N) fall inside the error bars. For the case of the GC l.o.s. clouds towards Sgr\,B2(N), their distances are uncertain and they are believed to be located within 1\,kpc from the GC (see Sect.\,\ref{sources}). As described in Sect.\,\ref{previous studies}, the purple and yellow hatched-shaded loci are derived from [Fe/H] vs $D$\textsubscript{GC} relations obtained by \citet{Bovy2014} and \citet{Genovali2014}, after accounting for the models of \citet{Kobayashi2011}; hatched-only loci correspond to an extrapolation of those relations, following \citet{Inno2019}. Using the same models, two $^{32}$S/$^{34}$S ratios are included. These ratios are derived from iron abundances measured in the central 30\,pc of the Galaxy by \citet{Davies2009} and \citet{Najarro2009}. A zoom on the results of our study displayed in the left part of the figure is shown in Fig.\,\ref{fig:zoom}.} 

  \label{fig:slopes}
\end{figure*}

Table 3 of \citet{Kobayashi2011} gives values for the $^{32}$S/$^{34}$S ratio as a function of [Fe/H] over the range \hbox{--0.5}$\leq$[Fe/H]$\leq$0.0. We can derive the following relation by interpolating their data:

\begin{equation}
\label{K1}
\frac{\rm{^{32}S}}{\rm{^{34}S}} = -19.8\times\rm{[Fe/H]}+23.2.
\end{equation}
Extrapolating up to [Fe/H]$\leq$0.85, we can account for the inner part of our Galaxy. Then, to relate $^{32}$S/$^{34}$S to the galactocentric distance, we can make use of the relations obtained by \citet{Bovy2014} and \citet{Genovali2014}, valid for 5$\leq$\textit{D}\textsubscript{GC}/kpc$\leq$14--19, respectively:
\begin{equation}
\label{Bovy}
\rm{[Fe/H]}=(-0.09\pm0.01)\times(\textit{D}\textsubscript{GC}/kpc-8)+0.03\pm0.01, 
\end{equation}
\begin{equation}
\label{Genovali}
\rm{[Fe/H]}=(-0.06\pm0.002)\times\textit{D}\textsubscript{GC}/kpc+0.57\pm0.02,    
\end{equation}
and combine them with Eq.\,\ref{K1}, to obtain the following relations:
\begin{equation}
\label{K1Bovy}
\frac{\rm{^{32}S}}{\rm{^{34}S}} = (1.8\mp0.2)\times(\textit{D}\textsubscript{GC}-8\,\rm{kpc})+22.6\mp0.2,    
\end{equation}
\begin{equation}
\label{K1Genovali}
\frac{\rm{^{32}S}}{\rm{^{34}S}} = (1.2\mp0.04)\times\textit{D}\textsubscript{GC}+11.9\mp0.4.
\end{equation}
Equations (\ref{K1Bovy}) and (\ref{K1Genovali}) are plotted in Fig.\,\ref{fig:slopes} as hatched-shaded regions. As described in the legend, their extrapolations down to 2.53\,kpc, following \citet[][their Figure 11]{Inno2019}, for both Eqs. \ref{Bovy} and \ref{Genovali}, are indicated as hatched regions only. 

Additionally, we have accounted for iron abundances obtained from high-resolution near-infrared observations by \citet{Davies2009} and \citet{Najarro2009} in the inner 30\,pc of the Galaxy \citep[see also][]{Kovtyukh2019}. Their measurements are [Fe/H]=0.1$\pm$0.2 and $-0.06$$\pm$0.2, respectively, and we converted them to $^{32}$S/$^{34}$S ratios of 21.2$\pm$4 and 24.4$\pm$4 by applying Eq.\,\ref{K1}.

In summary, the relations of both \citet{Bovy2014} and \citet{Genovali2014}, through Eq.\,\ref{K1}, give results closer to those obtained from the double-isotope method (Eq.\,\ref{eq:4}), considering $^{12}$C/$^{13}$C ratios by \citet{Yan2019} in combination with the $^{13}$C$^{32}$S/C$^{34}$S ratios from \citet{Chin1996}, as can be seen in Fig.\,\ref{fig:slopes} (dotted blue line). In the nuclear region of the Galaxy the \citet{Davies2009} and \citet{Najarro2009} observations, when accounting for Eq.\,\ref{K1}, are both consistent with our measurements for the +50\,km\,s$^{-1}$\,Cloud and the envelope of Sgr\,B2(N).

\section{Discussion}
\label{discussion}
Among the four stable sulphur isotopes ($^{32}$S, $^{33}$S, $^{34}$S, and $^{36}$S), $^{32}$S is a primary nucleus which could be synthesised in a single generation of massive stars. $^{32}$S is mostly formed during stages of hydrostatic and explosive oxygen-burning \citep{Wilson1992} either preceding a Type II supernova event or in a Type Ia supernova, where two $^{16}$O nuclei collide to form $^{28}$Si and $^{4}$He, with these products subsequently fusing to yield $^{32}$S. Type II supernovae synthesise around ten times more $^{32}$S than Type I supernovae, and occur roughly 5 times as often as those of Type I \citep{Hughes2008}. $^{33}$S is partly a secondary isotope because it can be formed by neutron capture from newly made $^{32}$S if the star not only has hydrogen and helium, but also carbon and oxygen in its initial composition \citep{Clayton2007}. It is synthesised in hydrostatic and explosive oxygen- and neon-burning, also produced in massive stars. $^{34}$S is partly a secondary product because it can be formed from newly made $^{32}$S and $^{33}$S by neutron capture, but also during oxygen burning in supernovae like the primary isotope, $^{32}$S \citep[and references therein]{Hughes2008}. While the comprehensive calculations of \citet{Woosley1995} identify $^{32}$S as a primary isotope, the same study also found that $^{34}$S is not a clean primary isotope; its yields decrease with decreasing metallicity. However, they identify $^{33}$S as a primary isotope, in contradiction with later findings \citep{Clayton2007}. $^{36}$S is probably the only purely secondary sulphur isotope, being produced by s-process nucleosynthesis in massive stars \citep{Thielemann1985,Mauersberger1996} and also by explosive C and He burning and via direct neutron capture from $^{34}$S, according to models \citep{Pignatari2016}. $^{36}$S could be the only S isotope not only produced from massive stars but also, to a lesser extent, from AGB stars \citep{Pignatari2016}. However, lines from C$^{36}$S are too weak to be detected in this study. Massive stars, as well as Type Ib/c and II supernovae, appear to slightly overproduce $^{34}$S and underproduce $^{33}$S compared to $^{32}$S, relative to the solar vicinity \citep{Timmes1995}. 
 
The main result of our study is that the previous trend observed by \citet{Chin1996} is broken near the centre of our Galaxy. In other words, the increase in $^{32}$S/$^{34}$S with $D$\textsubscript{GC} is not valid in the Galactic centre region. The values of 16.3$^{+2.1}_{-1.7}$ from the $+$50\,km\,s$^{-1}$ Cloud and 16.3$\pm$3.8 and 17.9$\pm$5.0 from the GC l.o.s. clouds towards Sgr\,B2(N) and its envelope, respectively, contrast with the expected $\sim$5--10 regardless of the value of $^{12}$C/$^{13}$C adopted (see Fig.\,\ref{fig:slopes}) when accounting for the $^{13}$C$^{32}$S/C$^{34}$S ratios used in \citet{Chin1996}. It is also worth mentioning that our $^{32}$S/$^{34}$S isotope ratios derived from absorption lines from diffuse or translucent clouds (Sgr\,B2(N)) are consistent with values derived from emission lines from a prominent star-forming region with dense molecular gas (the $+$50\,km\,s$^{-1}$\,Cloud), even though the chemistry for CS formation is completely different in those regions (see Appendix\,\ref{chemicalaspects}). 

\citet{Frerking1980}, even before the $^{32}$S/$^{34}$S slope was found, suggested values for $^{32}$S/$^{34}$S of $\sim$22 for the Galactic centre. Therefore, the sulphur ratio seems to be constant, or even increases with decreasing \textit{D}\textsubscript{GC}, within the inner 2.9\,kpc of the Milky Way, in contrast to $^{12}$C/$^{13}$C \citep[e.g.][]{Yan2019}, $^{14}$N/$^{15}$N \citep[e.g.][]{Adande2012}, and $^{18}$O/$^{17}$O \citep{Wouterloot2008, Zhang2015}. Intriguingly, $^{32}$S/$^{34}$S behaves in a similar way to $^{16}$O/$^{18}$O \citep{Polehampton2005}, two nuclei with the bulk of their formation taking place in massive stars ($\geq$10\,M$_{\odot}$) \citep{Clayton2007}. This is surprising because $^{34}$S is a tracer of secondary processing as $^{13}$C and $^{15}$N, and therefore its abundance is expected to increase in the same manner as observed for those isotopes.
 
The fact that sulphur traces late evolutionary stages of massive stars can give a clue to this difference in comparison to C and N, which give information on CNO and helium burning \citep{Chin1996}. Due to their short lives, the star formation rate of massive stars can be traced by their SN rate. Although the amount of $^{34}$S is related to metallicity, which decreases with increasing galactocentric radius especially in spiral galaxies \citep[e.g. as observed from oxygen,][]{Henry1999}, leading to a trend similar to C and N, the production of $^{34}$S is mostly related to SNe\,II, which show a dip in the inner regions of our Galaxy and other spiral galaxies \citep{Anderson2009}, in good agreement with our higher than expected $^{32}$S/$^{34}$S ratios in the Galactic centre. However, these results are still under debate \citep[see e.g.][]{Hakobyan2009} and more observations are needed. 

Another argument in favour of the above could be that metallicities traced by iron \citep{Genovali2014,Kovtyukh2019} instead of oxygen \citep{Henry1999} show a trend in good agreement with our observations (after converting [Fe/H] to \textit{D}\textsubscript{GC}/kpc not only outside the central 2.53\,kpc as in Fig.\,\ref{fig:slopes}, but also in the GC region itself). This could also indicate a drop in the production of massive stars at the Galactic centre compared to the rest of the Galaxy and to less massive stars.

Our $^{32}$S/$^{34}$S isotope ratio of 16.3$^{+2.1}_{-1.7}$ can constrain opacity estimates of sulphur-bearing molecules in the Galactic centre region and will considerably augment the confidence in theoretical modelling of dense molecular clouds \citep[e.g.][]{Loison2019}. Such results can then be used as initial inputs to reproduce hot core conditions \citep{Charnley1997,Viti2004,Vidal2018}.

Our new value for this ratio can also improve synthetic spectral fitting and subsequent line identification, giving better estimations to sulphur-bearing molecules. As a prime example, we mention the recent work of \citet{Zakharenko2019}: the canonical ratio of $^{32}$S/$^{34}$S = 22.5 \citep{Frerking1980} is insufficient to reproduce their data, as can be seen in their Fig.\,4. Our value of 16.3$^{+2.1}_{-1.7}$ leads to an enhancement of the $^{34}$S isotopologue by a factor of 1.4 in their fit (red line), better reproducing the lines detected at 99.512 and 99.520\,GHz, and thus increasing the confidence in the identification of CH$_{3}$$^{34}$SH in the hot core Sgr\,B2(N2).

\section{Summary and conclusions}
\label{summaryandconclusions}

From our analysis of the emission line profiles of CS, $^{13}$CS, C$^{34}$S, $^{13}$C$^{34}$S \textit{J}=2--1, of CS and C$^{34}$S \textit{J}=3--2, and of $^{13}$CS and C$^{34}$S \textit{J}=6--5 in the $+$50km\,s$^{-1}$ Cloud and CS, $^{13}$CS, C$^{34}$S, $^{13}$C$^{34}$S \textit{J}=2--1 observed in absorption towards Sgr\,B2(N), we obtain the following main results:
\begin{itemize}

\item From measurements of the \textit{J}=2--1 lines of $^{12}$C$^{34}$S \textit{J}=2--1 and $^{13}$C$^{34}$S \textit{J}=2--1, we have obtained a $^{12}$C/$^{13}$C isotope ratio of 22.1$^{+3.3}_{-2.4}$ near the centre of our Galaxy, in good agreement with previous estimations.

\item For the $+$50\,km\,s$^{-1}$\,Cloud we obtain a $^{34}$S/$^{33}$S ratio of 4.3$\pm$0.2, derived from C$^{34}$S \textit{J}=2--1 and C$^{33}$S \textit{J}=2--1 column densities. If we take the integrated intensities instead, this ratio would be 4.2$\pm$0.2, consistent with the lower end of the range of ratios obtained by \citet{Chin1996}, who derived $^{34}$S/$^{33}$S ratios between 4.38 and 7.53, irrespective of Galactic radius. This might be a first indication of a gradient with rising ratios as a function of increasing galactocentric distance, but data from the Galactic disc have to become more precise for a definite result.

\item From the \textit{J}=2--1 $^{13}$CS and $^{13}$C$^{34}$S emission lines in the $+$50\,km\,s$^{-1}$ Cloud, we derive, for the first time in a direct way, a $^{32}$S/$^{34}$S column density ratio of 16.3$^{+2.1}_{-1.7}$, which is consistent with the $^{32}$S/$^{34}$S ratio derived from the \textit{J}=6--5 and \textit{J}=2--1 $^{13}$CS and C$^{34}$S isotopologues when accounting for the above-mentioned $^{12}$C/$^{13}$C ratio. Due to possible CS fractionation, the above ratio might be underestimated by less than $\sim$10\%.

\item We were able to directly obtain a $^{32}$S/$^{34}$S ratio of 17.9$\pm$5.0 for the envelope of Sgr\,B2(N), from the isotopologues $^{13}$CS and $^{13}$C$^{34}$S in the \textit{J}=2--1 lines. Moreover, we have obtained a $^{32}$S/$^{34}$S ratio of 16.3$\pm$3.8 for the GC l.o.s. clouds towards Sgr\,B2(N) through the CS and C$^{34}$S \textit{J}=2--1 isotopologue lines, when CS is not optically thick. Those ratios are prone to increase by up to $\sim$15\%, when taking CS fractionation effects into account. 

\item Making use of the network presented in \citet{Loison2019}, we significantly improved CS fractionation estimations under conditions similar to those taking place in massive molecular clouds, young stellar objects, and diffuse or translucent cold molecular clouds. 

\item Comparing the sulphur ratios from this work with data available in the literature that were obtained from larger distances to Sgr\,A$^{\ast}$ and showed a decrease in $^{32}$S/$^{34}$S towards the Galactic centre, we can confidently establish that this decrease terminates at least at a distance of 100\,pc to Sgr\,A$^{\ast}$, at the position of Sgr\,B2(N) \citep{Reid2009}. This is different from trends previously reported for $^{12}$C/$^{13}$C, $^{14}$N/$^{15}$N, and $^{18}$O/$^{17}$O.

\item Our improved $^{32}$S/$^{34}$S isotope ratio will considerably augment the confidence in theoretical modelling for hot cores and in synthetic spectral fitting and subsequent line identification, giving better constraints for the intensities of sulphur-bearing molecules. 

\end{itemize}

Overall, our results suggest that processes occurring at late evolutionary stages of massive stars could be better traced by sulphur isotopologues instead of the most commonly studied CNO isotopes. Further observations targeting isotopologue ratios with distinct nucleogenesis like $^{32}$S/$^{34}$S (i.e. primary species versus secondary isotopologues) produced in advanced massive stars and SNe\,II can lead to a better understanding of environmental discrepancies between the solar neighbourhood and the inner Galaxy. This will allow, for example, a connection between  metallicity gradients traced by iron \citep{Genovali2014,Kovtyukh2019} and observations of SNe\,II \citep{Anderson2009}. The above could be also extrapolated to external galaxies, especially with the advent of a new generation of facilities.

\begin{acknowledgements}
We appreciate the important suggestions by the anonymous referee that helped to express this research in a clearer and more comprehensive way. PH is a member of and received financial support for this research from the International Max Planck Research School (IMPRS) for Astronomy and Astrophysics at the Universities of Bonn and Cologne. This work makes use of the following ALMA data: ADS/JAO.ALMA\#2011.0.00017.S, ADS/JAO.ALMA\#2012.1.00012.S. ALMA is a partnership of ESO (representing its member states), NSF (USA) and NINS (Japan), together with NRC (Canada), NSC and ASIAA (Taiwan), and KASI (Republic of Korea), in cooperation with the Republic of Chile. The Joint ALMA Observatory is operated by ESO, AUI/NRAO and NAOJ. The interferometric data are available in the ALMA archive at https://almascience.eso.org/aq/. This paper is partly based on data acquired with the Atacama PathfinderEXperiment (APEX). APEX is a collaboration between the Max Planck Institute for Radio Astronomy, the European Southern Observatory, and the Onsala Space Observatory. Part of this work is based on observations carried out with the IRAM 30-meter telescope. IRAM is supported by INSU/CNRS (France), MPG (Germany) and IGN (Spain). We acknowledge the IRAM staff for the support and help offered during all the applicable observational runs.
\end{acknowledgements}

\bibliographystyle{aa} 
\bibliography{biblio} 

\begin{appendix}
\section{Chemical aspects}
\label{chemicalaspects}
 
In diffuse clouds, it has been suggested that CS forms mainly by exothermic ion-molecule reactions of S$^{+}$ with CH and C$_{2}$ \citep{Drdla1989,vanDishoeck1998}. However, these species require to be one order of magnitude more abundant than currently observed in order to reproduce the observed quantities of CS \citep{Lucas2002}. Currently, CS formation is believed to be dominated by the exothermic reaction \citep{Drdla1989,Lucas2002,Montaigne2005,Laas2019}
\begin{equation}
\label{eqn:diffuse}
\rm{HCS^{+}+e^{-} \rightarrow CS + H}.
\end{equation}

\noindent
Subsequently, the dominant destruction route of CS is photodissociation, with destruction by He$^{+}$ being only significant in denser clouds \citep{Drdla1989}.

At higher densities, CS is mainly produced through neutral-neutral reactions with atomic sulphur such as \citep{Fuente2016,Vidal2018,Laas2019}
\begin{subequations}
\begin{equation}\label{eqn:1a}
\rm{S + CH \rightarrow CS + H},
\end{equation}
\begin{equation}\label{eqn:1b}
\rm{S + C_{2} \rightarrow CS + C}, and
\end{equation}
\begin{equation}\label{eqn:1c}
\rm{SO + C \rightarrow CS + O}.
\end{equation}
\end{subequations}

The dissociative recombination (DR) of HCS$^{+}$ becomes the main path for CS destruction at low temperatures because HCS$^{+}$ is mainly produced through protonation of CS and its DR leads mainly to S$+$CH and not to CS$+$H \citep{Montaigne2005}.

At higher temperatures, such as in hot cores and hot corinos for which recent models suggest some changes in sulphur chemistry \citep{Vidal2018}, CS is also destroyed by atomic oxygen:
\begin{equation}
\label{eqn:2}
\rm{CS + O \rightarrow CO + S}.
\end{equation}

\noindent
This reaction is supposed to be negligible at low temperature due to the presence of a barrier \citep{Lilenfeld1977,Gonzalez1996}. However the value of the barrier is questionable \citep{Adriaens2010} so that this reaction may not be completely negligible at low temperature. Apart from this reaction, and despite the recent advances in the investigation of sulphur species in the interstellar medium \citep{Fuente2016,Vidal2018,Laas2019}, there are still large uncertainties with respect to CS and HCS$^{+}$ chemistry.

Recent studies agree with the above reactions, but indicate that the dominant mechanism also depends on the cosmic ray ionisation rates (CRIR). At high CRIR, which may be typical of the central parts of giant spiral galaxies, reactions (\ref{eqn:1c}) and (\ref{eqn:2}) dominate CS formation and destruction, but at lower CRIR the following reaction (route\,(\ref{eqn:3})) becomes dominant for the destruction of CS \citep{Kelly2015,Viti2016}:
\begin{equation}
\label{eqn:3}
\rm{H_{3}O^{+} + CS \rightarrow HCS^{+} + H_{2}O}.
\end{equation}
In hot cores and corinos, recent models suggest some deviations from the above-mentioned routes of sulphur chemistry \citep{Vidal2018}. Specifically for CS, the main paths for its formation and destruction continue being Eq.\,\ref{eqn:1c} and Eq.\,\ref{eqn:2}, respectively. CS is initially destroyed by atomic oxygen (Eq.\,\ref{eqn:2}) both at 100 and 300\,K, on timescales of 10$^{4}$ and 10$^{3}$\,yr, respectively. Nevertheless, CS is then also produced by
\begin{equation}
\label{eqn:hotcores1}
\rm{S + CH_{2} \rightarrow HCS \overset{S}{\rightarrow} CS_{2} \overset{H}{\rightarrow} CS}.
\end{equation}

\section{Excitation temperatures}
\label{excitation_temperatures}

As mentioned in Sections \ref{peak_opacities_and_line_intesity_ratios} and \ref{csandc34sopacities}, we assumed the same excitation temperatures ($T_{ex}$) in all our opacity calculations. This is argued by the low opacity of the isotopologues used to make such calculations. 

\begin{figure}[h!]
\begin{minipage}{.9\linewidth}
\includegraphics[width=1.2\linewidth, trim = 0.cm 0.cm 0.cm 0.cm, clip=True]{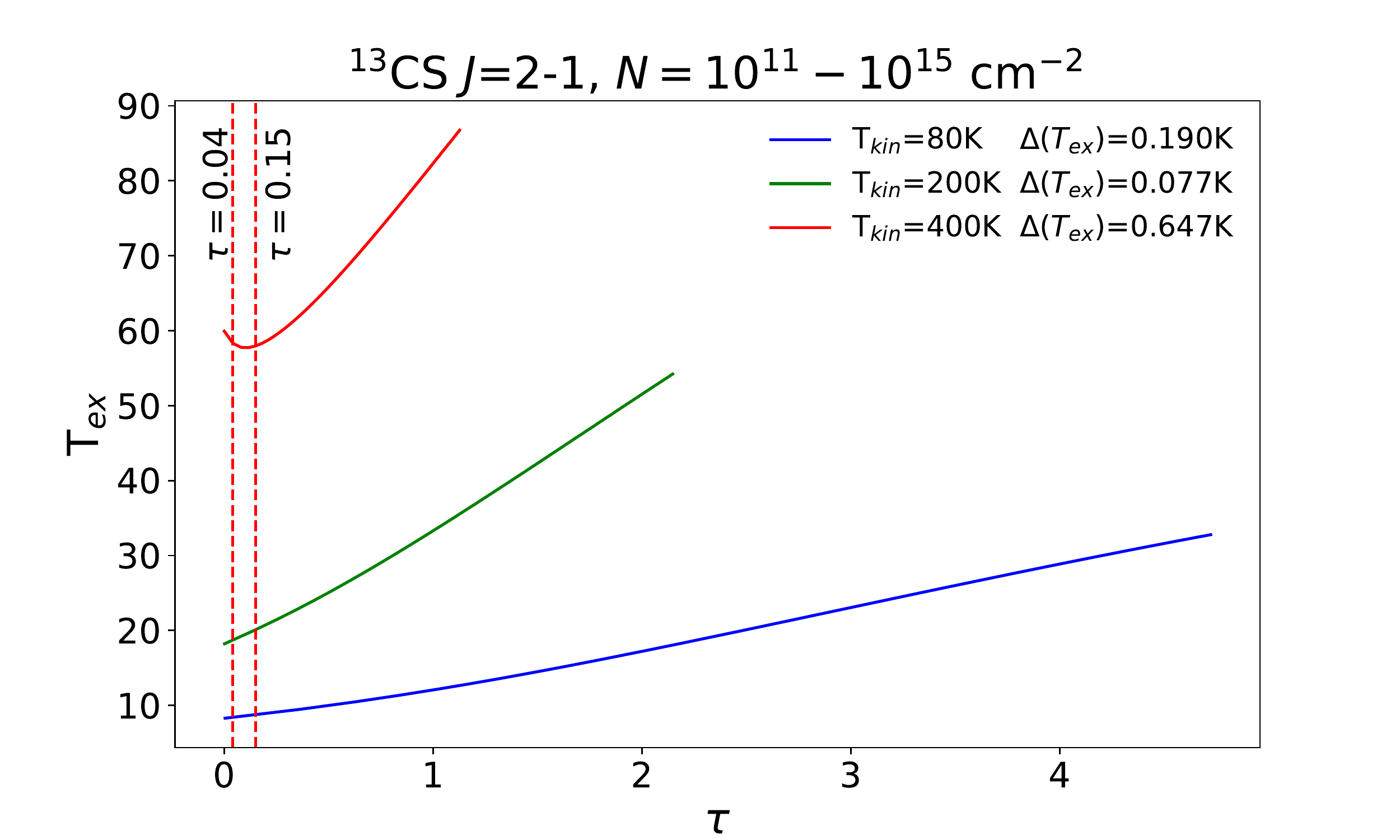}
\end{minipage}
\begin{minipage}{.9\linewidth}
\includegraphics[width=1.2\linewidth, trim = 0.cm 0.cm 0.cm 0.cm, clip=True]{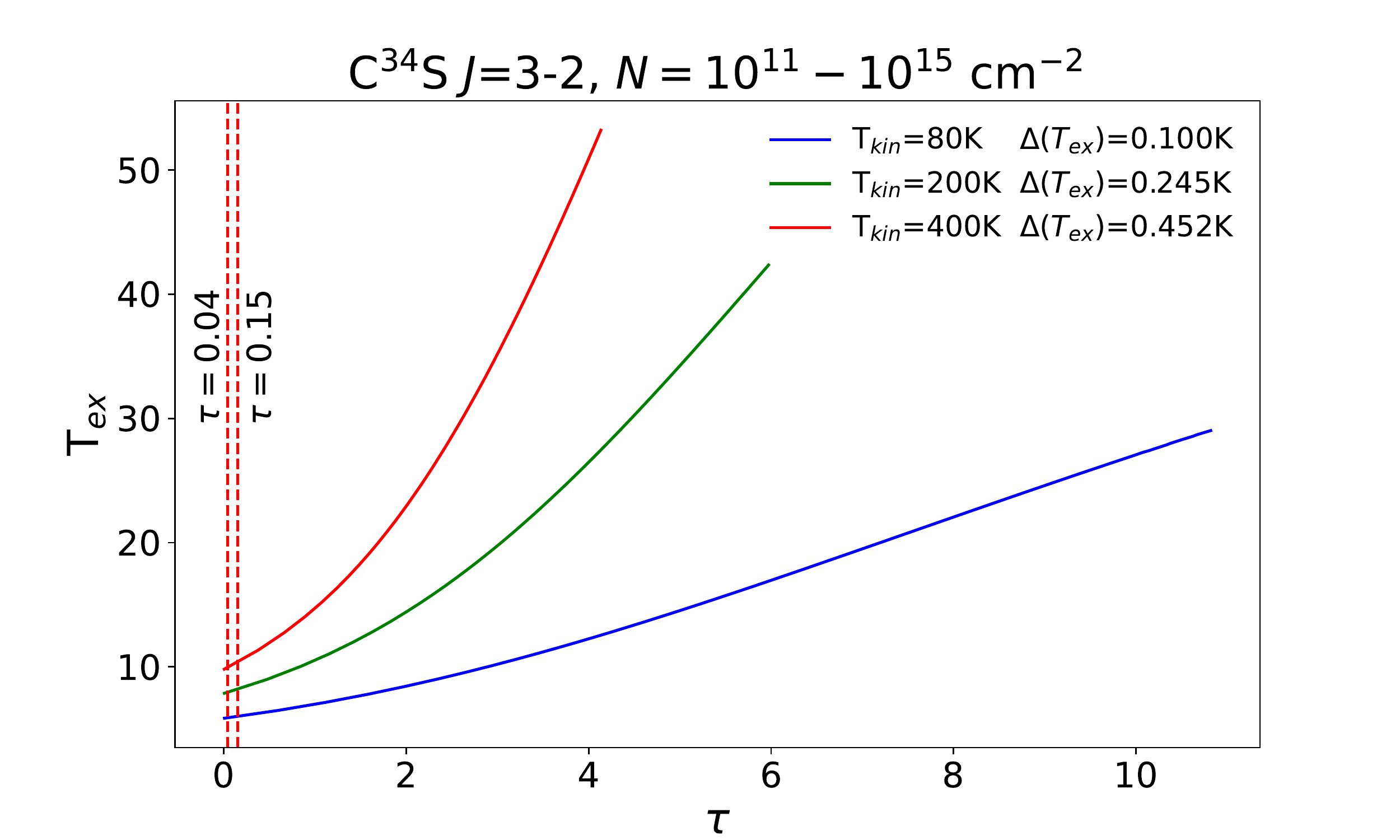}
\end{minipage}%
\caption{Excitation temperatures ($T_{ex}$) as a function of opacity ($\tau$) for $^{13}$CS \textit{J}=2--1 (top) and C$^{34}$S \textit{J}=3--2 (bottom). Molecular column densities range between 10$^{11}$ and 10$^{15}$\,cm$^{-2}$, and the kinetic temperature is 80, 200, and 400\,K, represented by blue, green, and red lines, respectively. $\Delta(T_{ex})$ refers to the $T_{ex}$ maximum variation between $\tau$=0.04 and 0.15 (delineated by dashed vertical red lines).}
\label{fig:excitation_temperatures}
\end{figure}

Let us take an opacity ($\tau$) ranging between 0.04 and 0.15 (see Table\,\ref{tab:2}) for both $^{13}$CS \textit{J}=2--1 and C$^{34}$S \textit{J}=3--2 with which we obtained the opacity for $^{12}$CS \textit{J}=2--1 and \textit{J}=3--2, respectively. As can be seen in Fig.\,\ref{fig:excitation_temperatures}, the variations in $T_{ex}$ within this range falls below 0.64\,K. Considering that this temperature will be inside an exponential as denominator of a fraction and, furthermore, multiplied by the Boltzmann constant \citep[see Eq.\,80 in ][]{Mangum2015}, is reasonable to assume that $T_{ex}$ variations are negligible for the scope of this work.

Calculations were done with a python version\footnote{see \url{https://pypi.org/project/ndradex/} for a full description. We have used the Grid RADEX run option.} of RADEX \citep{vanderTak2007}, a statistical equilibrium molecular radiative transfer code. Our inputs were a kinetic temperature of 80, 200, and 400\,K; 100 equally spaced values for the column density from 10$^{11}$ to 10$^{15}$\,cm$^{-2}$; and a H$_{2}$ volumetric density of 5$\times$10$^{4}$\,cm$^{-3}$, in an attempt to emulate the conditions of the $+$50\,km\,s$^{-1}$\,Cloud (see Section\,\ref{sources}). $^{13}$CS and C$^{34}$S collision rates were taken from the Leiden Atomic and Molecular Database \citep[LAMDA,][]{Schwoerer2019}, which in turn uses calculations made by \citet{Lique2006}.
 
\section{Zoom in for Figure 5}
\label{zoom}

We plot our own data from Fig.\,\ref{fig:slopes} in a magnified way. Data values for the l.o.s. clouds towards Sgr\,B2(N) and its envelope are given in Table\,\ref{isotopic_ratios_emoca}. The integrated intensity $^{32}$S/$^{34}$S ratio of the $+$50\,km\,s$^{-1}$\,Cloud (18.6$^{+2.2}_{-1.8}$) comes from Sect.\,\ref{ratiofromdirectobservations}.

\begin{figure}
  \includegraphics[width=\linewidth]{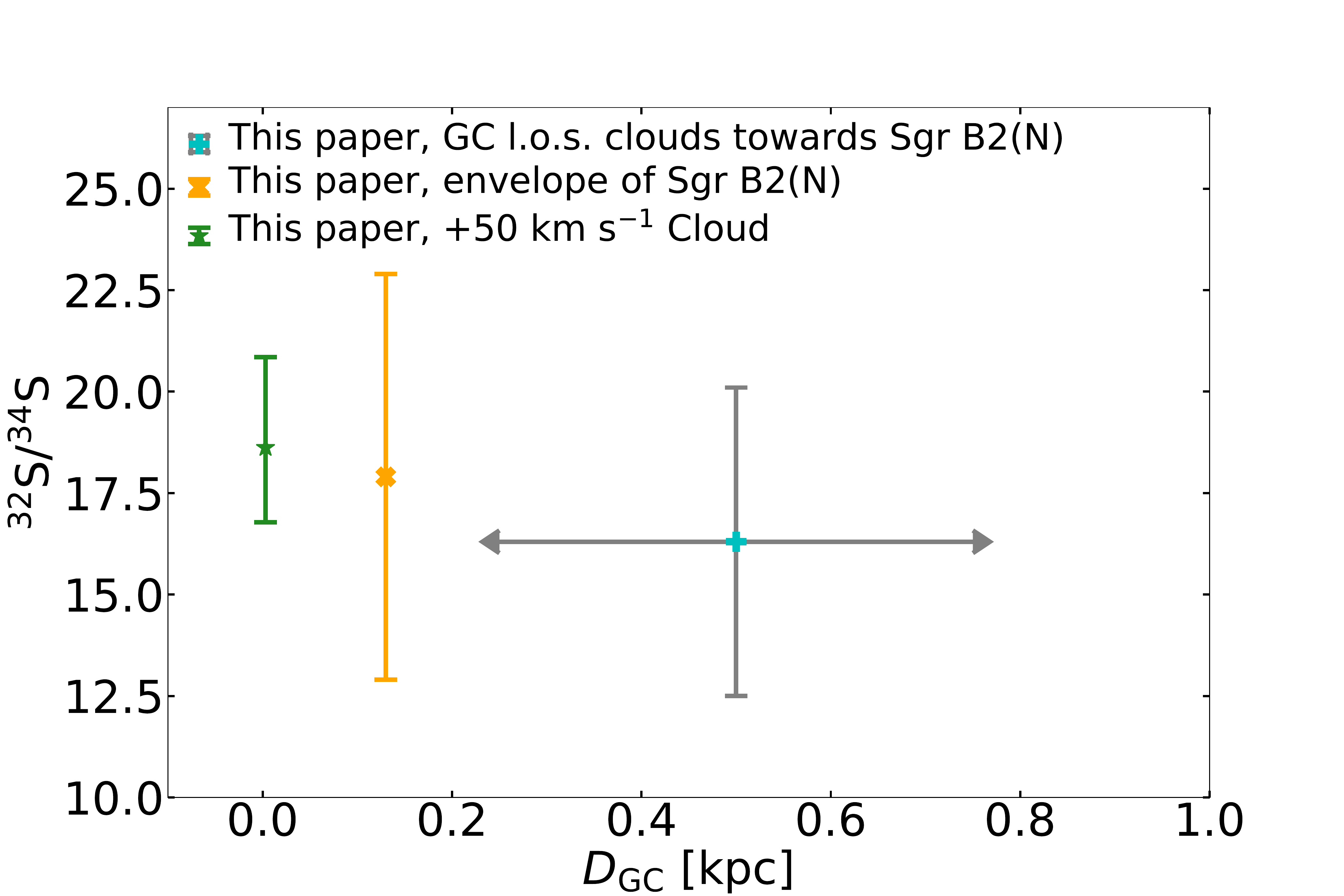}
  \caption{Sulphur isotope $^{32}$S/$^{34}$S ratio for the new data presented in this paper as a function of galactocentric radius, $D$\textsubscript{GC}.} 

  \label{fig:zoom}
\end{figure}

\section{Gaussian fitting}
\label{Gaussianfitting}

We preferred to use the lmfit package instead of Gildas\footnote{\url{https://www.iram.fr/IRAMFR/GILDAS}}, for example, because the former has some improvements related to uncertainty calculations. While both use minimisation, Simplex and Gradient methods, because of the use of the Minimize package\footnote{\url{https://lmfit.github.io/lmfit-py/fitting.html\#the-minimize-function}}, lmfit has additional functions and packages to ensure a proper uncertainty estimation 'even in the most difficult cases'\footnote{\url{https://lmfit.github.io/lmfit-py/}}. We compared the uncertainties from Gildas and lmfit; those derived based on Gildas are lower than those from lmfit by a small percentage (well within the uncertainties), and at the same time the use of Gildas results in a less successful fit in certain cases (not shown). Given the rich available documentation for lmfit\footnote{\url{https://lmfit.github.io/lmfit-py/}}, we opted for the lmfit-based fitting and associated uncertainty determination.

\end{appendix}

\end{document}